\documentclass[11pt]{article}
\usepackage[colorlinks=true,pdfstartview=FitV,linkcolor=blue,citecolor=blue,urlcolor=blue]{hyperref}
\usepackage{times}
\usepackage{amsmath,fullpage,empheq}
\usepackage{amsfonts}
\usepackage{amssymb}
\usepackage{graphicx}
\usepackage{enumerate}
\usepackage{multicol}  
\usepackage{relsize}
\usepackage[english]{babel}
\usepackage[utf8]{inputenc}
\usepackage{amsmath}
\usepackage[colorinlistoftodos]{todonotes}
\usepackage[super]{nth}
\usepackage{ amssymb }
\usepackage{blkarray}
\usepackage{tikz}
\usepackage{tikz-qtree}
\usepackage{subcaption}
\usepackage{url}
\usepackage{diagbox}
\usepackage{mathtools}

\DeclarePairedDelimiter\floor{\lfloor}{\rfloor}
\usepackage{amsthm}
\usepackage{verbatim}
\usepackage[ruled,vlined]{algorithm2e}
\SetKw{KwBy}{by}
\theoremstyle{definition}

\newtheorem{definition}{Definition}

\newtheorem{theorem}{Theorem}[section]

\author{Andrew Richards and Laura Kubatko}
\title{Bayesian Weighted Triplet and Quartet Methods for Species Tree Inference}
\begin{document}
\maketitle

\section{Abstract}
Inference of the evolutionary histories of species, commonly represented by a species tree, is complicated by the divergent evolutionary history of different parts of the genome. Different loci on the genome can have different histories from the underlying species tree (and each other) due to processes such as incomplete lineage sorting (ILS), gene duplication and loss, and horizontal gene transfer. The multispecies coalescent is a commonly used model for performing inference on species and gene trees in the presence of ILS. This paper introduces Lily-T and Lily-Q, two new methods for species tree inference under the multispecies coalescent. We then compare them to two frequently used methods, SVDQuartets and ASTRAL, using simulated and empirical data. Both methods generally showed improvement over SVDQuartets, and Lily-Q was superior to Lily-T for most simulation settings. The comparison to ASTRAL was more mixed -- Lily-Q tended to be better than ASTRAL when the length of recombination-free loci was short, when the coalescent population parameter $\theta$ was small, or when the internal branch lengths were longer.

\section{Introduction}
\begin{figure}
            \centering
        \includegraphics[width=0.8\linewidth]{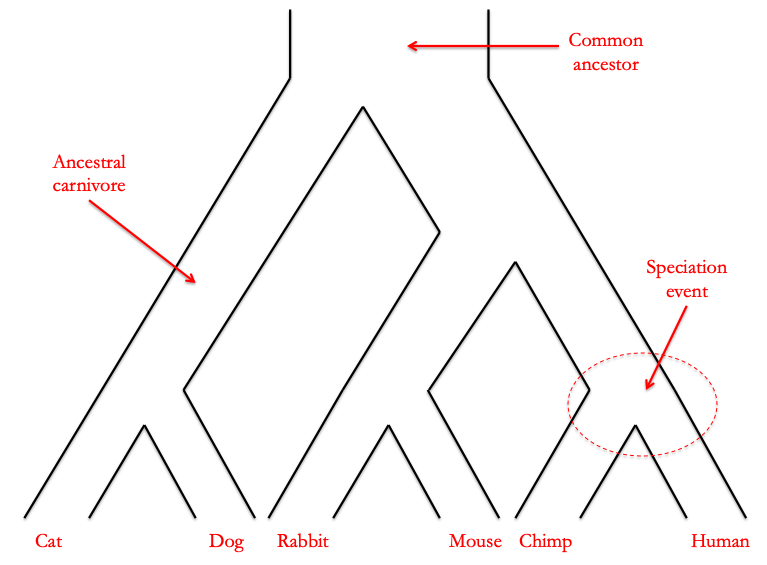}
    \caption{Example phylogenetic tree featuring six mammalian species.}
    \label{exampleTree}
\end{figure}

The phylogenetic inference problem is concerned with using data, including but not limited to DNA sequences, to understand the evolutionary history of a collection of species. Consider the collection of mammals shown in figure \ref{exampleTree}. We are concerned with three aspects shown in the figure. First, is the unlabeled topology correct? In other words, for each node, how many descendants are there on the left and right branches? Second, is the labeled topology correct, or does it need to be permuted? Third, when do the speciation events occur?

We usually begin with the assumption that when a speciation event occurs each ancestral species divides into exactly two daughter species. Thus, the evolutionary history can be represented as a binary tree known as a species tree.
\begin{definition}
A \textit{species tree} is an acyclic graph $S=(V(S), E(S), \boldsymbol{\tau}_S)$ where $V(S)$ is the vertex set of $S$, $E(S)$ is the edge set of $S$, and $\boldsymbol{\tau}_S$ is a set of branch lengths. 
\end{definition}
\label{Species tree}
Biologically, internal nodes represent speciation events while the leaves represent extant species. We call this leaf set $L_S$. The leaves will be represented by lower case letters $a, b, c,$ etc., and internal nodes by letters later in the alphabet. Internal branches represent ancestral species. If we know the common ancestor of all the species under consideration, then the tree is \textit{rooted}, and the tree becomes a directed graph from the root outward. If the root is unknown, then we only know the direction of the branches that connect to external nodes. The \textit{degree} of a node is the number of other nodes a node is connected to. For a species tree, the leaves are of degree one, and due to the binary tree assumption, all internal nodes are of degree three (except the root, if known, which is of degree two).

A common and useful assumption, known as the \textit{molecular clock}, is that the mutation rate is constant over time (or more precisely, if $\lambda$ represents the mutation rate there is a common $\lambda(t)$ for all branches at time $t$). In that case, the rooted tree is \textit{ultrametric}, meaning that each leaf will be equidistant from the root. For ultrametric trees, an equivalent parameterization to the set of branch lengths $\boldsymbol{\tau}_S$ is the set of node times $\boldsymbol{\tau}_{S_{nodes}}$. Because we assume the molecular clock, we will simplify the notation so that $\boldsymbol{\tau}$ represents the node times for the remainder of the paper.

\begin{figure}[h]
    \centering
    \includegraphics[width=0.8\linewidth]{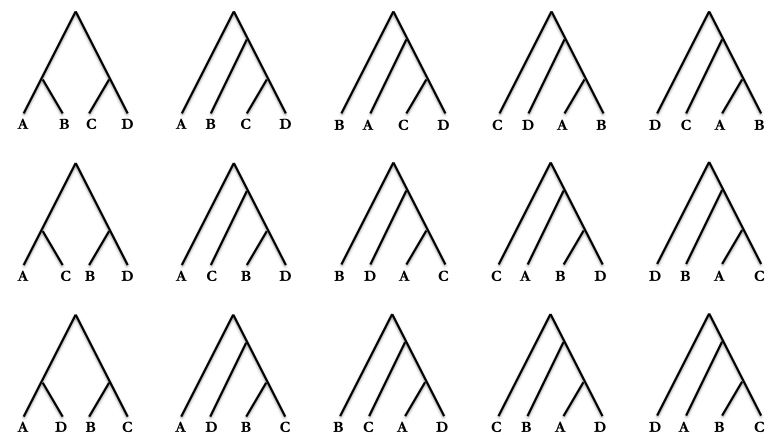}
    \caption{The fifteen rooted quartets. We can label these $\mathcal{S}^{(4)}_1, \mathcal{S}^{(4)}_2 \ldots \mathcal{S}^{(4)}_{15}$. If we are unconcerned with the labels, the first column is $\mathcal{S}^{(4)}_{u1}$ and the other columns $\mathcal{S}^{(4)}_{u2}$. Note that each row corresponds to one of the three unrooted quartets.}
    \label{quartets}
\end{figure}

Let $\mathcal{S}^{(n)}$ refer to the set of all possible trees with $n$ taxa. The superscript is in parentheses to highlight that it refers to the number of taxa rather than as an exponent. Then, for example, $\mathcal{S}^{(4)}_1$ and $\mathcal{S}^{(4)}_2$ can indicate the first and second 4-taxon species trees from figure \ref{quartets}. If the labeling is unimportant, we can also use $\mathcal{S}^{(4)}_{u1}$ and $\mathcal{S}^{(4)}_{u2}$ to indicate the two \textit{unlabeled} 4-taxon topologies. 

Species tree inference is complicated by the possibility for divergence between the evolutionary history of species and individual elements of their genomes. Causes for this divergence include incomplete lineage sorting (ILS), gene duplication and loss (GDL) and horizontal gene transfer (HGT). ILS is commonly modeled by the coalescent process \cite{kingman1982}. The history of individual loci on the genome is represented by a gene tree.

\begin{definition}
A \textit{gene tree} is a network $G=(V(G), E(G), \boldsymbol{t}_G)$ where $V(G)$ is the vertex set of $G$, $E(G)$ is the edge set of $G$, and $\boldsymbol{t}_G$ is a set of branch lengths. When the molecular clock is assumed, $\boldsymbol{t}_G$ can equivalently represent node times as with the species tree above and the notation simplified to $\boldsymbol{t}$. These node times are subject to the constraint that the coalescent events in question must occur prior to the divergence time of the species in question.
\end{definition}

\begin{figure}[h]
\centering
\begin{subfigure}[h]{.48\linewidth}
   \centering
\includegraphics[width=.9\linewidth]{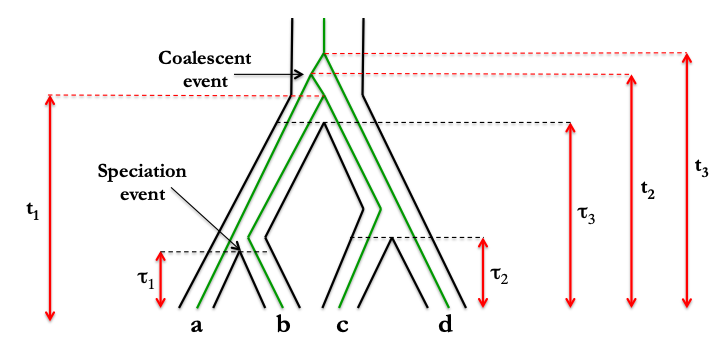}
\caption{Symmetric species tree}
\end{subfigure}
\begin{subfigure}[h]{.44\linewidth}
\centering
\includegraphics[width=.9\linewidth]{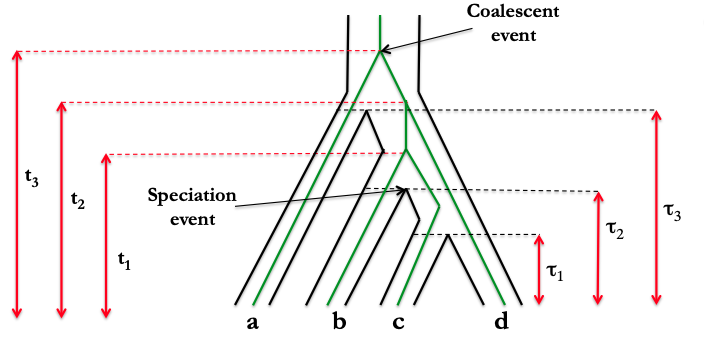}
\caption{Asymmetric species tree}
\end{subfigure}
\caption{Species trees with four taxa under the coalescent process. The green lines show example gene trees evolving within the underlying species tree.}
\label{coalescent}
\end{figure}

The gene tree is embedded within the species tree (see figure \ref{coalescent}) and usually both trees have the same leaf set. Exceptions can occur if the gene has not been sampled for all species under consideration or if there are multiple sampled individuals per species. Thus, unless otherwise noted, we will drop the subscript and just refer to the set of species under consideration as $L$. When we need to distinguish the leaves of $G$ from the leaves of $S$ we use capital letters $A, B, C, \ldots$. Care should be taken, however, to distinguish when $A$ or $C$ are used to denote members of the leaf set and when they are used as abbreviations for nucleotides. Internal nodes represent coalescent events, which identify the most recent common ancestor of two gene lineages. 

It is worth noting here that our definition of a gene, as is common in the phylogenetic literature, refers to a recombination-free region of the genome. Thus, the common assumption is that no recombination occurs within a gene, and all sites in a gene share a common evolutionary history, while sites on separate genes are independent conditional only on the underlying species tree. This differs from the biological definition of a gene as a segment of DNA coding for a polypeptide. There is, of course, no underlying reason why a biological gene can or should share a common history in its entirety. To avoid this confusion, \textit{locus} is also sometimes used to describe a recombination-free region, however, it is still more common to refer to ``gene trees" rather than ``locus trees".

A number of different approaches have been taken with regard to species tree inference in the presence of ILS. The first is essentially to ignore the problem: perform gene tree inference on concatenated data using methods such as RAxML \cite{raml} or FastTree \cite{FastTree}, treating all sites as if they share a single, common evolutionary history. This can be fast and accurate for estimating $S$. But, there are some concerns: concatenation has been shown to be statistically inconsistent for some values of $(S,\boldsymbol{\tau})$ \cite{degnan05,roch15}, and speciation time estimates are biased since the coalescent event must naturally occur before the speciation time. Another approach is the use of summary statistics that first estimate the gene trees independently for each gene, and then use the gene tree estimates as inputs for species tree estimates. Examples of this approach include STEM \cite{kubatko09}, ASTRAL \cite{mirarab14,mirarab15,zhang18}, and MP-EST \cite{MPEST}. These methods can be computationally efficient, however they depend on the accuracy of the gene tree estimates that are used as inputs as well as proper delineation of recombination-free segments of the genome \cite{gatesy14,springer16}. A third approach uses the full data to coestimate the species tree and each of the gene trees, generally using Markov chain Monte Carlo (MCMC) methods. Examples of this approach include BEST \cite{BEST}, *BEAST \cite{BEAST,SB2},  and BPP \cite{BPP}. These methods can be quite accurate but are very computationally intensive when the number of loci and/or leaf set is large. Assessment of convergence is also a challenge, especially due to the multi-modal nature of the likelihood in the tree space \cite{salter01}.

A fourth approach, and the one we take in this paper, is to treat the gene trees as a nuisance parameter that can be integrated over. A previous example of this approach is SVDQuartets \cite{chifman14}, which uses a rank-based methodology to infer the proper unrooted species tree for each quartet of species under consideration, and then uses an assembly algorithm to infer the final n-taxon species tree estimate taking the set of unrooted quartet trees as input. The theory behind the method assumes unlinked coalescent-independent sites (CIS) data such that the gene tree underlying each site can be treated as a random draw from the distribution of all possible gene trees given the species tree, however Wascher and Kubatko (2020)\nocite{wascher20} recently proved that SVDQuartets is statistically consistent for multilocus data as well.

Our method, \textbf{(Li)}kelihood-based assemb\textbf{(ly)} (\textbf{Lily}), also assumes unlinked CIS data. From \cite{chifman15} we have the site pattern probabilities given a species tree topology and branching times. Then, a prior on these topologies and branching times is assumed, and posterior probabilities for each set of rooted triplets (Lily-T) or unrooted quartets (Lily-Q) are calculated. These posterior probabilities are then used as weights in an assembly algorithm to infer the final n-taxon estimated topology $\hat{S}$. The details of the procedures are described in the next section.

\section{Method}
The outline for the Lily-T and Lily-Q procedures are laid out in algorithms \ref{tProd} and \ref{Qprod}. For each triplet (Lily-T) or quartet (Lily-Q) of species, first the site pattern frequencies are found. Then, the likelihood for each rooted topology is calculated. Using Bayes's Theorem, the posterior probability of each rooted triplet or unrooted quartet is then determined. Finally, given these posteriors as inputs, the n-taxon topology is estimated using supertree assembly methods. Each step is discussed in detail in the following sections.
\begin{algorithm}
\SetAlgoLined
  \For{$l\gets1$ \KwTo ${n \choose 3}$}{
    Find site pattern frequencies $\boldsymbol{D}_{JC}$ for the $l$\textsuperscript{th} triplet of species (section \ref{dataSec})\;
    Find the site pattern probabilities $\boldsymbol{\delta}|(S,\boldsymbol{\tau)}$ for each of the three rooted triplet topologies from \cite{chifman15} (section \ref{method1} and appendix \ref{tripletApp})\;
    Integrate over $\boldsymbol{\tau}$ using $\theta=0.003$ and $\beta$ as estimated from equation \ref{distEst} to find $\boldsymbol{\delta}|S$ and then $L(S|\boldsymbol{D}_{JC})$ for each rooted triplet (sections \ref{derivation} and \ref{robustness})\;
    From Bayes's Theorem find the posterior probability of each of the three rooted triplets (section \ref{derivation})
    }
    Using the $3{n \choose 3}$ posterior probabilities as input, estimate the n-taxon rooted topology using the Triplet MaxCut algorithm \cite{sevillya16} (section \ref{assembly})\;
 \caption{Lily-T procedure}
 \label{tProd}
\end{algorithm}

\begin{algorithm}
\SetAlgoLined
  \For{$l\gets1$ \KwTo ${n \choose 4}$}{
    Find site pattern frequencies $\boldsymbol{D}_{JC}$ for the $l$\textsuperscript{th} quartet of species (section \ref{dataSec})\;
    Find the site pattern probabilities $\boldsymbol{\delta}|(S,\boldsymbol{\tau)}$ for each of the fifteen rooted quartet topologies from \cite{chifman15} (section \ref{method1})\;
    Integrate over $\boldsymbol{\tau}$ using $\theta=0.003$ and $\beta$ as estimated from equation \ref{distEst} to find $\boldsymbol{\delta}|S$ and then $L(S|\boldsymbol{D}_{JC})$ for each rooted quartet (sections \ref{derivation} and \ref{robustness})\;
    From Bayes's Theorem find the posterior probability of each of the fifteen rooted quartets (section \ref{derivation}) \;
    Calculate the posterior probability of the three unrooted quartet topologies as the sum of the corresponding rooted quartets (see figure \ref{rooting})\;
    }
    Given the $3{n \choose 4}$ posterior probabilities as input, estimate the n-taxon rooted topology using the Weighted QMC algorithm \cite{avni14} (section \ref{assembly})\;
 \caption{Lily-Q procedure}
 \label{Qprod}
\end{algorithm}

\subsection{Data structure}
\label{dataSec}
\begin{figure}[h]
    \centering
    \includegraphics[width=0.8\linewidth]{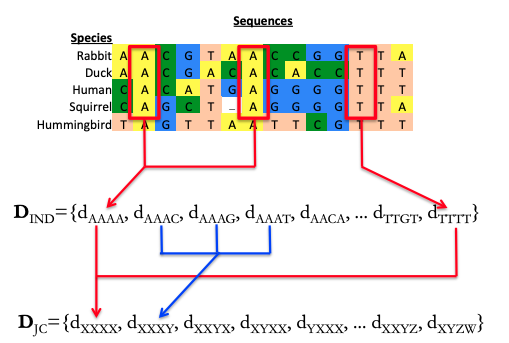}
    \caption{Data reduction (Lily-Q): For each set of four species, the raw aligned sequence data can be reduced down to $\boldsymbol{D}_{ind}$ and then $\boldsymbol{D}_{JC}$ under the Jukes-Cantor assumptions.}
    \label{data_map}
\end{figure}

Our data structure and data reduction method for Lily-Q are summarized in figure \ref{data_map}. We begin by assuming a matrix of aligned sequence data $\textbf{D}_{raw}$ where the rows represent the species under consideration and the columns each represent an aligned site. We assume in the sequel that this alignment has been performed without error. Even with a correct alignment, there may also be sites present in one sequence and not another, either due to the data truly being missing or because of an insertion or deletion. Thus, $D_{ij} \in \{A,C,G,T,-\}$.  Here, $i$ is an index for the taxon, $j$ is the index for the site, the letters represent the four nucleotides, and the dash represents missing data.

If the data are iid, or if the possibility of varying rates across sites is treated as a random effect that can be integrated over, then the columns are exchangeable. We begin with one subset of four of the $n$ taxa. Let $\delta_{TCCG}$ represent the probability that at a certain site the first species has A, the next two have C, and the fourth species has G, i.e., $\delta_{TCCG}=P(i_A=T, i_B=C, i_C=C, i_D=G)$ where $i_x$ is the state for species $x$. As discussed before, the abbreviation C is overloaded so care should be taken to note that $i_C=C$ means the third species has cytosine at this site. Under this assumption of exchangeability, $\delta_{TCCG}$ will be the same at every site. Then, the number of sites where the first taxon has character $i_A$, the second has character $i_B$, etc., which we label $d_{i_A, i_B, \ldots i_n}$ will follow a binomial distribution with probability $\delta_{i_A,i_B, \ldots i_n}$, and the joint probability of all possible site patterns is a multinomial distribution. Since there are $4^n$ possibilities (or $5^n$ with missing data), the numbers of sites that follow each pattern $d_{i_A, i_B, \ldots i_n}$ is a sufficient statistic. Then we can map down $\boldsymbol{D}_{raw}$ down to a $4^n \times 1$ vector $\boldsymbol{D}_{ind} \sim Multinom(J,\boldsymbol{\delta}_{ind})$, where each element of $\boldsymbol{\delta}$ represents one of the site pattern probabilities and $J$ is the total number of sites. In the sequel, we will only consider sites where all four species in the quartet have a nucleotide present -- i.e., sites where $D_{ij} \in \{A,C,G,T\}$, $i=A,B,C,D$, which will not affect inference if sites are missing at random. 

If we further assume the Jukes-Cantor (JC69) substitution model \cite{jukes69}, all nucleotides have the same limiting frequency of 1/4 and all substitution rates between nucleotides are the same. As a result, for determining probabilities, if we use the JC69 model we don't need to keep track of what nucleotides are present where; we only need to note whether or not they are the same. For example with two taxa, $P(i_A=C, i_B=C)=P(i_A=G, i_B=G)$, as is the probability of any two different nucleotides at the same site. So, we can call two identical nucleotides at the same site an XX pattern regardless of whether they represent AA, CC, GG, or TT. Similarly, XY represents the case where the two nucleotides are different. Again, the distribution of the site pattern frequencies follows a multinomial distribution, but we need to keep track of fewer cases. The number of different cases required for $n$ taxa is $\sum_{i=1}^4 S_{n,i}$ where $S_{n,i}$ is the Sterling number of the second kind. Most relevant for our later discussion, for a four-taxon tree, we can map $\boldsymbol{D}_{ind}$ down to a $15 \times 1$ vector $\boldsymbol{D}_{JC}$. We repeat this mapping for each quartet of species, creating an ${n \choose 4}$ set of site pattern frequency vectors $\boldsymbol{D}_{JC}$.

This data mapping process is similar for Lily-T, except that it is repeated of each of the ${n \choose 3}$ set of triplets. $\boldsymbol{D}_{ind}$ then is a $4^3 \times 1$ vector and  $\boldsymbol{D}_{JC}$ is a $5 \times 1$ vector.

\subsection{Derivation of $L((S,\boldsymbol{\tau)|\boldsymbol{D}_{JC}})$ for 3- or 4-taxon trees}
\label{method1}
Chifman and Kubatko (2015)\nocite{chifman15} derived the site pattern probabilities $\boldsymbol{\delta}|(S,\boldsymbol{\tau})$ for a 4-taxon tree where $\delta_k|(S,\boldsymbol{\tau})$  is the probability of the $k$\textsuperscript{th} site pattern, $k \in \{XXXX, XXXY, \ldots XYZW\}$, occurring at a given site given the species tree topology and branching times under the JC69 model and the molecular clock. Then since $\boldsymbol{D}_{JC} \sim Multinom(J, \boldsymbol{\delta}|(S,\boldsymbol{\tau}))$ the likelihood is given by 

\begin{equation}
    L((S,\boldsymbol{\tau})|\boldsymbol{D}_{JC}) \propto \prod_{k=1}^{K} [\delta_k|(S,\boldsymbol{\tau})]^{d_k} 
    \label{lik1}
\end{equation}

To find $P(\boldsymbol{D}_{JC}=\boldsymbol{d_{JC}}|(S,\boldsymbol{\tau}))$ we first recognize that this can be factored into two processes: the coalescent process and the substitution process:
$$ P(\boldsymbol{D}_{JC}=\boldsymbol{d_{JC}}|(S,\boldsymbol{\tau}))=P(\boldsymbol{D}_{JC}=\boldsymbol{d_{JC}}|(G,\boldsymbol{t}),(S, \boldsymbol{\tau}))f((G,\boldsymbol{t})|(S, \boldsymbol{\tau})),  $$
$$=P(\boldsymbol{D}_{JC}=\boldsymbol{d_{JC}}|(G,\boldsymbol{t}))f((G,\boldsymbol{t})|(S, \boldsymbol{\tau})).  $$
where the second equality is true under the assumption of neutral selection, whereby the substitution process and coalescent process are independent and then the first term depends directly only on the gene tree \cite{whidden15}.

We begin by noting that:
$$\delta_{XXXY}|((a,(b,(c,d))),\boldsymbol{\tau}))=\delta_{XYXX}|((a,(d,(c,b))),\boldsymbol{\tau})).$$ A similar argument can be made for all fifteen site pattern frequencies and all fifteen rooted 4-taxon topologies. Thus, we need only derive the site pattern probabilities for the two topologies shown in figure \ref{coalescent}, $((a,b),(c,d))$ and $(a,(b,(c,d)))$ and the site pattern probabilities for the other 13 topologies is a permutation of one of these two cases. The likelihood of any other 4-taxon tree is then given by equation \ref{lik1} with the data permuted as necessary by the permutation function $\sigma(\cdot)$:
\begin{equation}
L((S,\boldsymbol{\tau})|\boldsymbol{D}_{JC}) \propto \prod_{k=1}^{K} [\delta_k|(S,\boldsymbol{\tau})]^{\sigma(d_k)} 
\label{lik2}
\end{equation}

$\boldsymbol{\delta}|(S,\boldsymbol{\tau})$ for the lone unlabeled 3-taxon tree topology can be found be marginalizing over the fourth taxa in figure \ref{coalescent}. The details are shown in the Appendix.

\subsection{Derivation of $P(S|\boldsymbol{D}_{JC})$ for 3- or 4-taxon trees}
\label{derivation}
From the Law of Total Probability, it is immediate that: 
$$ \boldsymbol{\delta}|S = \int_{\boldsymbol{\tau}} \boldsymbol{\delta}|(S,\boldsymbol{\tau}) f(\boldsymbol{\tau}) d\boldsymbol{\tau}$$
A wide variety of forms for the density $f(\boldsymbol{\tau})$ can be chosen. We choose priors to be uninformative and to allow for analytic solutions to $\boldsymbol{\delta}|S$. Refer to figure \ref{coalescent} for a description of $\boldsymbol{\tau}$ and note that the 3-taxon case can be viewed as the asymmetric case with taxon $a$ removed. For a 3-taxon tree, we choose $\tau_2 \sim Exp(\beta)$ and $\tau_1|\tau_2 \sim U(0,\tau_2)$. For the 4-taxon symmetric tree, we choose $\tau_3 \sim Exp(\beta)$ and $\tau_1|\tau_3,\tau_2|\tau_3$ independently $\sim U(0,\tau_3)$ For the 4-taxon asymmetric tree, we choose $\tau_3 \sim Exp(\beta)$ and $(\frac{\tau_1}{\tau_3}, \frac{\tau_2}{\tau_3}) \sim Dirichlet(1,1,1)$. Choosing a prior on the root age gives us a prior on the age of the tree as a whole; setting exponential priors on each branch length leads to the total age of the tree being dependent on the degree of tree symmetry. Given these priors on the branching times, $\boldsymbol{\delta}|S$ is calculated as derived in the appendix and the likelihood of any tree can be found as before by taking these site pattern probabilities and applying the standard multinomial likelihood to a permutation of the data:
$$L(S|\boldsymbol{D}_{JC}) \propto \prod_{k=1}^{K} [\delta_k|S]^{\sigma(d_k)} $$

An unfortunate side effect of this choice of prior is that for four-taxon trees (by a simple application of the law of conditional expectation) $E(\tau_1)=\frac{1}{2}\tau_3$ for the symmetric topology and $E(\tau_1)=\frac{1}{3}\tau_3$ for the asymmetric topology. In other words, if (C,D) is a cherry, inferring an asymmetric topology automatically implies the prior assumption that this divergence time between species C and D occurs more recently. This does not appear to be a problem at first glance, since we are not concerned with inferring species divergence times here, and are treating $\boldsymbol{\tau} $ as a nuisance parameter. But, it turns out that this makes accurate inference of the root location of a 4-taxon tree impossible. Given our choice of model, if the true tree is symmetric -- for example $((a,b),(c,d))$ -- then various pairs of rooted asymmetric trees have the same likelihood with speciation times integrated out: $ L[(a,(b,(c,d)))]=L[(b,(a,(c,d)))]$, $ L[(c,(d,(a,b)))]=L[(d,(c,(a,b)))]$, etc. Worse, for some true values of $\tau_1, \tau_2, \tau_3$, $ L[(a,(b,(c,d)))]>L[((a,b),(c,d))]$. These inference errors all concern the location of the root rather than the unrooted topology. As a result, we limit our inference on 4-taxon trees to unrooted topologies. 

For the prior on topologies, we assume the tree generation follows a Yule model: there is a constant rate of species divergence over time, and the rate of species divergence is equal for all branches. This is meant to be as uninformative as possible. For 3-taxon trees, this intuitively assigns a 1/3 probability to each of the 3 rooted 3-taxon trees. For 4-taxon trees, each of the three symmetric topologies has a 1/9 prior probability and each of the twelve asymmetric topologies has a 1/18 prior probability. Interestingly, while any individual symmetric topology is twice as probable as any individual asymmetric topology, in the prior it is twice as probable that the unlabeled topology will be asymmetric rather than symmetric. The details of the prior calculation are given by \cite{harding71}.

\begin{figure}[h]
    \centering
    \includegraphics[width=0.6\linewidth]{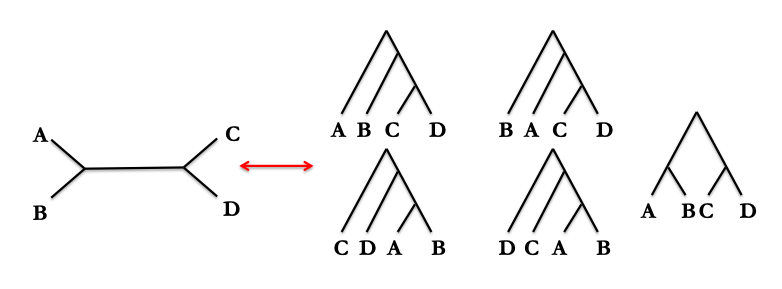}
    \caption{Each unrooted 4-taxon tree corresponds to five different rooted trees, arising from placing the root on one of the five branches in the unrooted tree.}
    \label{rooting}
\end{figure}

Taking both the prior on the topologies and each topology's likelihood, it is a simple application of Bayes's Theorem in the 3-taxon case to show:
$$ P(S=s|\textbf{D})=\frac{L(S|\textbf{D})P(S=s)}{\sum_{i=1}^3 L(S_i|\textbf{D})P(S_i=s_i)} $$ 
For the 4-taxon case, the summation is performed over 15 rather than 3 topologies, and the final probability of the unrooted topology is the sum of the five rooted topologies compatible with it (see figure \ref{rooting}).

For the 3-taxon case, we can show that given a sufficient number of unlinked sites under the molecular clock, we can infer the correct topology with probability 1. For the unrooted 4-taxon case, we have demonstrated this in simulation studies for a wide variety of true trees, but it remains unproven.
\begin{theorem}
Given CIS data for three taxa evolving under the multispecies coalescent with the JC69 model and the molecular clock, if $\hat{S}$ is the maximum a posteriori tree, $P(\hat{S}=S) \rightarrow 1$ as the number of sites $J \rightarrow \infty$ for any prior for which $0<\tau_1<\tau_2$ holds with probability 1.
\end{theorem}
Then, using a Bonferroni adjustment, we can ensure that given a sufficient number of sites, $P(\hat{S}=S) > 1- \epsilon$ \textit{uniformly} for all the triplets. It then follows that given proposition four from \cite{steel93}, we can estimate the n-taxon topology with probability $>1-\epsilon$.

\subsection{Estimating the n-taxon species tree}
\label{assembly}
There is no theoretical impediment to prevent extending this procedure to infer full posterior probabilities for 5-, 6-, or even n-taxon trees. But, to see the practical difficulties, consider the challenges to extend this result to just five taxa. First, for 5 taxa $\boldsymbol{\delta}_{JC}$ is now a $51 \times 1$ vector for each of three unlabeled topologies. Then, a $51 \times 105$ permutation matrix is needed to find the site pattern probabilities for all 105 5-taxon labeled gene tree topologies. A general recursive formula for the number of unordered gene histories embedded in an n-taxon species tree is provided by \cite{rosenberg07}. We calculated this total to be 379, 313, and 208 for $S_1$, $S_2$, and $S_3$ from figure \ref{topologies}, respectively. The probabilities of these histories need to be calculated and then integrated over to find the site pattern probabilities of the three unlabeled species tree topologies, then the same $51 \times 105$ permutation matrix must again be applied to get the full site pattern probabilities for all species tree topologies. As a result, the full set of site pattern probabilities $\boldsymbol{\delta}|(S,\boldsymbol{\tau})$ remains to be calculated for 5 taxa, and an even more complex process would be needed to extend to six or more taxa.

\begin{figure}
    \centering
    \includegraphics[width=0.8\linewidth]{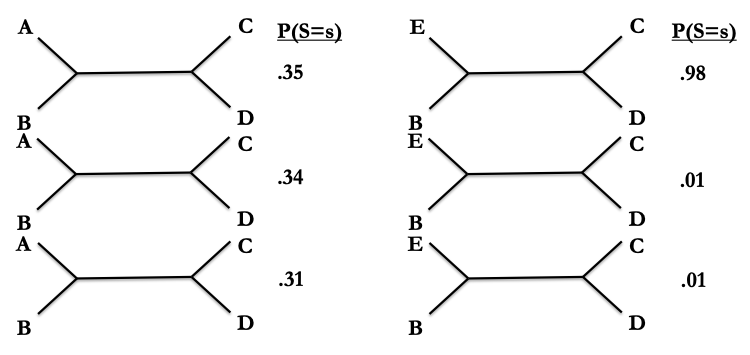}
    \caption{Unweighted tree inputs: an unweighted method treats the two cases the same, even though the right side contains far more information about the true topology.}
    \label{weighting}
\end{figure}

Instead, to infer the n-taxon topology, we use an assembly algorithm that takes the posterior probabilities of the rooted triplets or unrooted quartets as inputs to infer the final n-taxon species tree estimate. We repeat the process of sections \ref{method1} and \ref{derivation} for each of the ${n \choose 3}$ sets of triplets (Lily-T) or ${n \choose 4}$ quartets (Lily-Q). Then we have a set of $3{n \choose 3}$ posterior probabilities for each possible rooted triplet or $3{n \choose 4}$ posterior probabilities for each possible unrooted quartet from the leaf set. These probabilities are then used in an assembly algorithm to infer the estimated n-taxon species tree $\hat{S}$. There are many options for assembly methods. The methods we choose allow us to use the posterior probabilities as weights in the subtree inputs. Both SVDQuartets and ASTRAL are unweighted methods and thus treat all inputs equally regardless of the inference uncertainty (see figure \ref{weighting}). For Lily-T, we chose the Triplet MaxCut (TMC) method of Sevillya et al. (2016)\nocite{sevillya16} due to its speed, accuracy, ease of implementation, and its ability to work with rooted trees as inputs, and similarily chose the Weighted Quartet Max Cut (weighted QMC) algorithm of Avni et al. (2015)\nocite{avni14} for Lily-Q. Further, the implementation of these algorithms are very similar to the algorithm used by SVDQuartets, and so using it reduces, but does not eliminate, one source of variation between the two methods.

Triplet MaxCut works as follows: first we obtain posterior probabilities on all ${n \choose 3}$ subsets from the leaf set. When we have multiple individuals from a species, posterior probabilities for each combination of individuals from each triplet of species are calculated. Then two symmetric $n \times n$ matrices are formed, $\textbf{G}$ and $\textbf{B}$, called the ``good" and ``bad" matrices with $i,j=1,2,3 \ldots n$ corresponding to the taxa in alphabetical order: $a, b, c,$ etc. For each triplet input, we have a set of two taxa that form a cherry and a third that is not part of the cherry, as well as a weighting for the triplet. For each of the ${3 \choose 2}$ pairs in the triplet, if taxa $i$ and $j$ form a cherry, the weight of the triplet is added to $B_{ij}$ and $B_{ji}$ and if $i$ and $j$ do not form a cherry, the weight of the triplet is added to $G_{ij}$ and $G_{ji}$. As an example, consider the triplet $(a,(c,d))$ with weight 0.5. We would add 0.5 to $B_{34}$, $B_{43}$, $G_{13}$, $G_{31}$, $G_{14}$, and $G_{41}$. This process is repeated for all sets of input triplets. Thus, large entries in $G_{ij}$ indicate taxa $i$ and $j$ should, all else equal, be separated, and large entries in $B_{ij}$ indicate taxa $i$ and $j$ should, all else equal, be grouped together in the final tree S.

Next, for the set of $n$ taxa $L=\{1,2,3 \ldots\ n\}$ we obtain all subsets of the taxa of cardinality $\leq \floor*{\frac{n}{2}}$. For each subset, we can arbitrarily label the taxa in the subset $L_1$ and call the remainder $L_2=L\setminus L_1$, resulting in a bipartitioning of $L$. Then, for each bipartition, we score the bipartition by the ratio $ \frac{\sum_{i \in L_1} \sum _{j \in L_2} G_{ij}}{\sum_{i \in L_1} \sum _{j \in L_2} B_{ij}}$. Note that this score can be undefined if there are entries in $\boldsymbol{B}=0$. To avoid this problem, a very small number such as $10^{-200}$ can be added to each element of $\boldsymbol{B}$. The bipartition with the highest score is accepted. If the cardinality of either $L_1$ or $L_2$ is greater than two, the process is repeated recursively on $L_1$ and/or $L_2$ as needed. In essence, each step of the procedure creates a node and assigns taxa to the left and right branches of the node until the final tree is resolved.

The operation of the weighted QMC is largely similar, with posteriors first being calculated for all ${n \choose 4}$ subsets of quartets from the leaf set to use as inputs in the assembly algorithm. The major difference in the assembly is that care must be taken if one of $L_1$ or $L_2$ is a set of 3 elements. Then the set is augmented with an additional artificial taxon and the procedure is performed on this augmented set to evaluate which two elements of the set constitute a cherry.

\subsection{Uncertainty quantification}
An immediate concern with using an assembly procedure is that one of the key advantages of our likelihood-based approach -- the ability to produce posterior probabilities -- is lost. The assembly procedures we chose are based on heuristics that make sense -- grouping species together that have a high probability of being cherries. But, unfortunately we have no distributional results to assess. The output tree is a point estimate only, and while we can generate simulation data to say that it is reasonably accurate, once we apply it to real data we no longer have a measure of uncertainty. The difficulty lies in the same factors that led us to pursue the assembly procedure in the first place -- the inability to calculate a joint $n$-taxon set of site pattern probabilities.

A standard method for measuring uncertainties of estimated phylogenies is the bootstrapping method of Efron (1979) \cite{efron79}. We implemented a nonparametric bootstrap: for CIS data we resampled with replacement from all sites, and for multi-locus data we resampled the genes with replacement and then for each gene we resampled with replacement the sites within each gene. An advantage of the bootstrap is that it can give unbiased estimates of uncertainty without any distributional assumptions. But, a number of cautions are in order. First, we only have asymptotic guarantees about the approximation of $\frac{\boldsymbol{d}_k}{J}$ to $\delta_k$ and we have no finite-sample knowledge of how good the approximation is. As a result, we don't know how much uncertainty we have about our uncertainty. Second, without parallelization, the time required to estimate the bootstrap samples increases linearly with the number of bootstrap samples taken. Last, bootstrap support values are \textit{not} probabilities, and should not be treated as such. That said, we can compare the bootstrap support to the actual proportion of times we infer the correct tree to see if the bootstrap support reasonably ``mimics" the true probability of being correct for reasonable parameter values.

\section{Implementation}
\label{implement}
We have written source code in C++ that implements Lily-T and Lily-Q, as well as programs that summarize the distances from the true tree for our simulation runs. The programs take as inputs alignments in PHYLIP format and output either the final output tree (Lily-T) or a properly formatted input file for the weighted QMC program (Lily-Q). We also wrote programs for calculating the number of gene tree histories in section \ref{assembly} and a Perl wrapper for executing the simulation runs. These programs, as well as a file summarizing the results are available at \url{https://github.com/richards-1227/Lily}. 

\section{Results}
Data were simulated using the \textit{ms} \cite{MS} and Seq-Gen \cite{seq-gen} programs in C++ using a Unix HPC platform. The \textit{ms} program takes the node times (in coalescent units) and population parameter $\theta$ as input and simulates a set of gene trees under the multispecies coalescent model. Seq-Gen then takes these gene trees and the mutation model parameters as input and generates the aligned sequence data $\boldsymbol{D}_{raw}$. Caution should be taken in working with these programs for diploid organisms -- to get the correct simulated probabilities, you enter one-half the node times (in coalescent units) and $2\theta$ rather than their actual values. 1,000,000 CIS were simulated at various settings of $\boldsymbol{\tau}$ and $\theta$ and we compared the observed values of $\boldsymbol{d}_{JC}$ to $J\boldsymbol{\delta}_{JC}|(S,\boldsymbol{\tau})$. Chi-squared goodness of fit tests were performed and there was no evidence that the simulated frequencies differed from that expected (only one individual test p-value was below 0.05 (0.006), and it was no longer significant after making a Bonferroni adjustment).

\subsection{Testing robustness to model assumptions and prior specification}
\label{robustness}
We next relaxed various assumptions of the substitution model to verify that the true tree was still estimated with high probability. One thousand runs were performed with 25,000 CIS generated with the Jukes-Cantor assumptions being progressively relaxed, first allowing for different stationary probabilities, then different relative substitution rates, then between-site rate heterogeneity, and then allowing for invariant sites. The settings are summarized in table \ref{JCsettings}.

\begin{table}[h]
    \centering
    \begin{tabular}{|c|c|c|c|c|c|c|c|c|c|}
    \hline
       &   & \multicolumn{5}{|c|}{Relative rates vs. A-C} &  &  & \\
      \hline
      Setting & CG content  & A-G & A-T  & C-G  & C-T  & G-T  & $\alpha$ & \% Invariable &  $P(S|\boldsymbol{d})$\\
       \hline
        1 & 25\% & 1 & 1 & 1 & 1 & 1 & N/A & 0 & 0.983      \\ 
               \hline
        2 & 35\% & 1 & 1 & 1 & 1 & 1 & N/A & 0   & 0.971\\ 
        \hline
                3 & 45\% & 3 & 1 & 1 & 3 & 1 & N/A & 0 & 0.951 \\ 
        \hline
                4 & 35\% & 5 & 1 & 1 & 3 & 1 & N/A & 0  & 0.951\\ 
        \hline
                5 & 35\% & 5 & 0.75 & 0.5 & 3 & 0.25 & N/A & 0 & 0.971 \\ 
        \hline
                6 & 35\% & 2 & 0.75 & 0.5 & 1.5 & 0.25 & 3 & 0 & 0.974  \\ 
        \hline
                7 & 35\% & 2 & 0.75 & 0.5 & 1.5 & 0.25 & 9 & 0  & 0.960\\ 
        \hline
                8 & 35\% & 2 & 0.75 & 0.5 & 1.5 & 0.25 & 3 & 0.2  & 0.960 \\ 
        \hline
    \end{tabular}
    \caption{Initial simulation test settings. The final column gives the average posterior probabilities assigned to the true tree as the JC69 assumptions are progressively relaxed.}
    \label{JCsettings}
\end{table}

The results shown in table \ref{JCsettings} are typical and are presented for three taxa and each of the eight substitution model settings using $\tau_1=5.2$, $\tau_2=6.0$, $\theta=0.01$, and $\beta=0.1$. There does not appear to be a large impact from relaxing the JC69 assumptions, and for the remainder of our simulations all the data were generated under simulation setting 8, so performance of our methods is measured against a deliberately misspecified model.

\begin{figure}
    \centering
   \begin{subfigure}[b]{0.43\textwidth}
   \includegraphics[width=\linewidth]{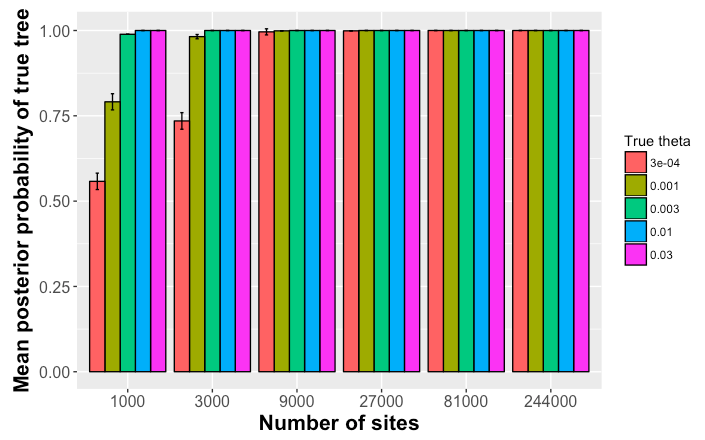}
   \caption{$\hat{\theta}=0.001$}
\end{subfigure}
\begin{subfigure}[b]{0.43\textwidth}
   \includegraphics[width=\linewidth]{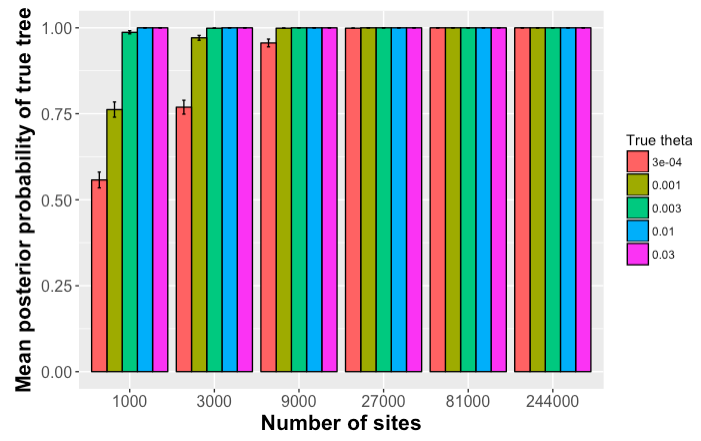}
   \caption{$\hat{\theta}=0.01$}
\end{subfigure}
    \caption{Mean posterior probability assigned by Lily-Q to the true tree for varying numbers of sites and population parameter $\theta$ for a symmetric 4-taxon  with branch lengths of 0.5 coalescent units. The left panel (a) uses $\hat{\theta}=0.001$ and the right panel (b) uses $\hat{\theta}=0.01$.}
    \label{ThetaEffect}
\end{figure}

Here we should note that the site pattern probabilities $\boldsymbol{\delta}$ are also conditional on $\theta$ as well as $(S,\boldsymbol{\tau})$, but keeping with the notation of \cite{chifman15} we have ignored this conditioning. For the remainder of our work, we used $\hat{\theta}=0.003$ as an input to both Lily-T and Lily-Q, which we justify as follows. Most empirical values of $\theta$ fall between 0.001 and 0.01 (see \cite{kubatko07,rannala03,jennings05,kopp05} for estimates on species ranging from primates to finches). So, we simulated data using a true $\theta=\{0.0003, 0.001, 0.003, 0.01, 0.03\}$, extending beyond the empirical range, and calculated posteriors using $\hat{\theta}=0.01$ and $\hat{\theta}=0.001$. The results for Lily-T and Lily-Q are shown in figure \ref{ThetaEffect}. Since there is no apparent difference in performance for assuming a $\theta$ different from the actual data-generating $\theta$, we simply use a $\hat{\theta}=0.003$ as it is in the middle of the empirical range on the log scale.

\begin{figure}
    \centering
    \includegraphics[width=0.9\linewidth]{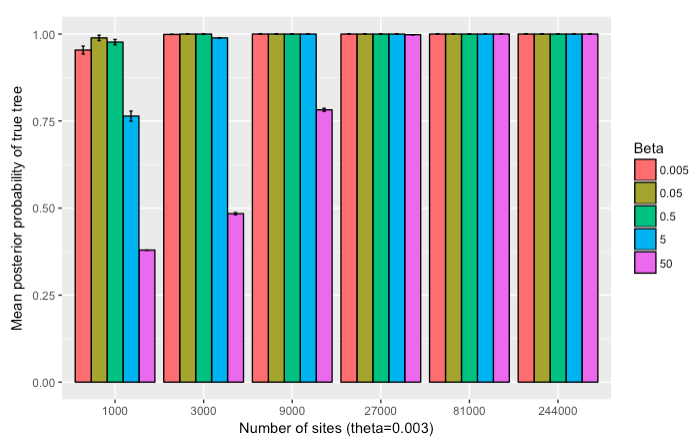}
    \caption{Mean posterior probability assigned by Lily-T to the true tree for varying numbers of sites and root age hyperparameter $\beta$ for a 3-taxon tree with branch lengths of 0.5 coalescent units.}
    \label{betaEffect}
\end{figure}

We cannot, however, simply assume a value for the root age hyperparameter $\beta$. Figure \ref{betaEffect} shows the posterior assigned to the correct tree, using a $\beta$ where $1/\beta$ varies from 2 orders of magnitude below to 2 orders of magnitude above the actual root age. There is very little loss in accuracy for overestimating the root age (very small values of $\beta$) except with very few sites (see the $J=1000$ column). If the root age is underestimated, however, there can be a large loss of accuracy, even with a larger number of sites. This is to be expected given the asymmetric nature of the exponential prior -- a small $\beta$ flattens the prior and can still have adequate prior mass near the true value, while a large $\beta$ puts most prior mass near zero. 

We take an empirical Bayes approach to choosing $\beta$. From the JC69 model we can infer the well-known pairwise distance estimate for each pair of species
$$\hat{\lambda} \hat{t}=\frac{-3log(1-\frac{4\hat{p}}{3})}{4} $$
where $t$ is the time measured in years, $\lambda$ is the per-year mutation rate, and $\hat{p}$ is the fraction of discordant sites between the two species. Then, assuming a value of 0.003 for $\theta$ and generation time $g$, we have:
        $$0.003=\theta=4N_e\mu=4\lambda N_e g $$
After a quick algebraic manipulation we can estimate time in coalescent units ($2N_e$ generations) as:
  \begin{equation}
      \frac{\hat{t}}{2N_eg}=\frac{-6log(1-\frac{4\hat{p}}{3})}{0.012} 
      \label{distEst}
  \end{equation}      
For each set of three or four species, we will have either three or six of these pairwise divergence estimates from equation \ref{distEst}. Lily-T and Lily-Q use the mean of these estimates as the value for the hyperparameter $\beta$. It is worth noting that because these estimates use concatenated data, they are biased upwards with respect to the species tree root time since they estimate the time of coalescent events and any coalescent event must occur prior to species divergence. But, here, that bias is useful as it helps reduce the risk of underestimating the root age, which has a greater performance impact than overestimation. 

\subsection{Application to simulated data}
\begin{figure}
    \centering
    \begin{subfigure}{.47\linewidth}
\centering
\includegraphics[width=\linewidth]{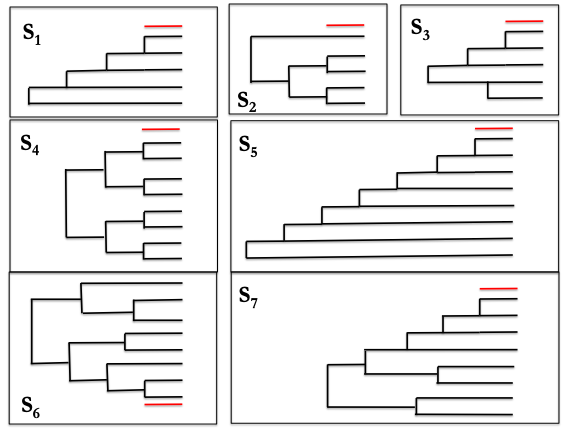}
\end{subfigure}
\begin{subfigure}{.47\linewidth}
\centering
\includegraphics[width=\linewidth]{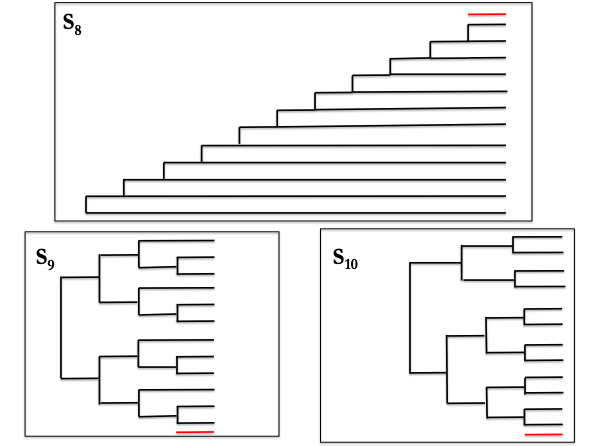}
\end{subfigure}
\newline
\begin{subfigure}{.47\linewidth}
\centering
\includegraphics[width=\linewidth]{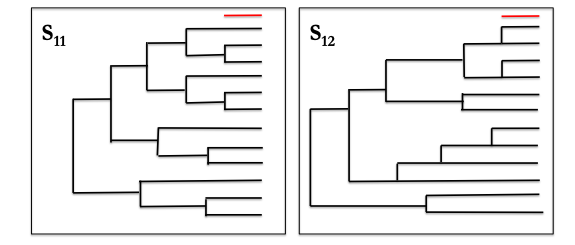}
\end{subfigure}
    \caption{The twelve topologies used in the simulation study. 5 taxa: $S_1$ through $S_3$, 8 taxa: $S_4$ through $S_7$, 12 taxa: $S_8$ though $S_{12}$. The red bar indicates scale of the MBL, which was set at 0.2, 0.5, 1.0, and 2.0 coalescent units.}
    \label{topologies}
\end{figure}

Data were simulated for 5, 8, and 12-taxon trees for a total of 12 different topologies ranging from fully symmetric to fully asymmetric. The full set of topologies simulated is displayed in figure \ref{topologies}. For each topology, we used four different values of the population parameter: $\theta=\{0.0008, 0.0024, 0.0072, 0.0216\}$. These were chosen to extend slightly above and below the empirical range of 0.001 to 0.01 and the values are linear on the log scale. We used four different settings for the minimum branch length (MBL): 0.2, 0.5, 1.0, and 2.0 coalescent units. Multi-locus data were simulated with 10, 50, and 500 genes and 50, 200, and 500 sites per gene. For CIS data, we simulated either 5,000 or 50,000 sites. 100 runs were performed at each combination of settings.

Focusing on the minimum branch length created a number of side effects. Comparing trees $S_4$ and $S_5$ in figure \ref{topologies} we note that for the same MBL, the root for the asymmetric tree $S_5$ will occur much further in the past than for the symmetric tree $S_4$. Second, for trees that are neither fully symmetric nor asymmetric such as $S_{12}$, some internal branches naturally have to be longer than the MBL, in which case the internal nodes were chosen to be equally spaced. 

In order to evaluate how well each method performed, we first need a proper metric of distance between trees. The most common is the Robinson Folds (RF) distance from \cite{RFdist}. They define the distance between two trees $T_1$ and $T_2$ as the sum of the number of bipartitions in $T_1$ not contained in $T_2$ and vice versa. The presence of a bipartition is symmetric, since the number of bipartitions is equal to the number of internal branches, which is in turn equal to $n-2$ for a rooted tree and $n-3$ for an unrooted tree. Thus, if there is a bipartition of $T_1$ not in $T_2$, there must be a bipartition in $T_2$ not in $T_1$. As a result, the RF metric can take on any even value from zero to $2(n-2)$ (or $2(n-3)$ if unrooted). Since the metric is also a function of whether or not the two trees are rooted, to compare Lily-T to the other methods (which produce unrooted trees), we must first unroot both the estimated and true tree. For multi-locus data, we compared SVDQuartets (as implemented in PAUP* \cite{paup}), ASTRAL (using FastTree for gene tree estimation as per the example in \cite{chou15}), Lily-T, and Lily-Q for each run and calculated the mean RF distance and standard error for the 100 runs. For CIS data, we excluded ASTRAL from comparison since gene tree cannot be estimated for a single site.

\begin{figure}
\centering
   \begin{subfigure}[b]{0.49\textwidth}
   \includegraphics[width=\linewidth]{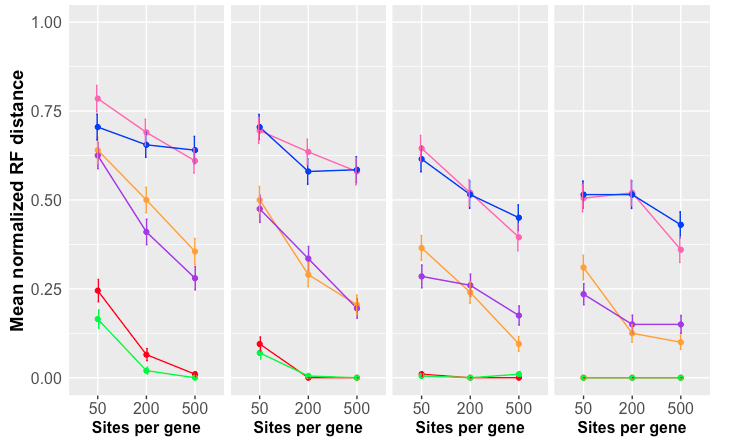}
   \caption{$MBL=0.2$}
\end{subfigure}
\begin{subfigure}[b]{0.49\textwidth}
   \includegraphics[width=\linewidth]{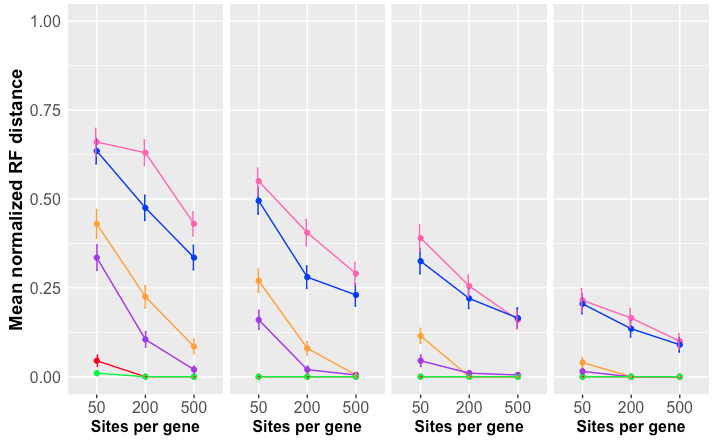}
   \caption{$MBL=0.5$}
\end{subfigure}
\newline
   \begin{subfigure}[b]{0.49\textwidth}
   \includegraphics[width=\linewidth]{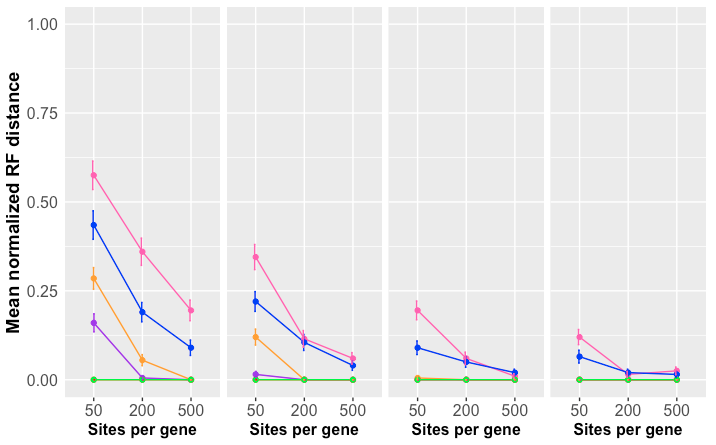}
   \caption{$MBL=1.0$}
\end{subfigure}
\begin{subfigure}[b]{0.49\textwidth}
   \includegraphics[width=\linewidth]{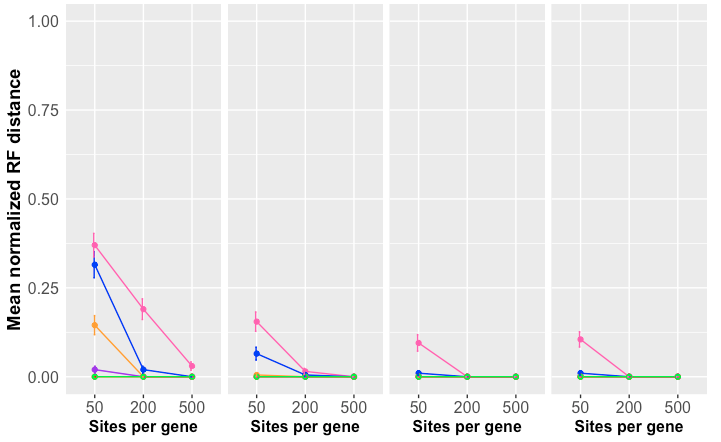}
   \caption{$MBL=2.0$}
\end{subfigure}
\caption{Mean normalized RF distance for Lily-Q vs. ASTRAL for true tree $(a,(b,(c,(d,e))))$ at different minimum branch lengths (MBL). In each panel, $\theta=\{0.0008, 0.0024, 0.0072, 0.0216\}$ from left to right. Number of genes: 10 -- magenta/blue, 50 -- orange/purple, 500 -- red/green. ASTRAL is displayed in ``hot" colors (magenta/orange/red) and Lily-Q in ``cold" (blue/purple/green). }
\label{simulation_RF1}
\end{figure}

\begin{figure}
\centering
   \begin{subfigure}[b]{0.49\textwidth}
   \includegraphics[width=\linewidth]{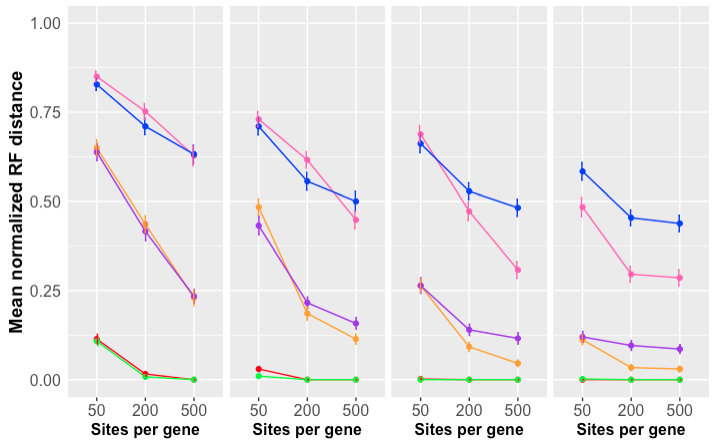}
   \caption{$MBL=0.2$}
\end{subfigure}
\begin{subfigure}[b]{0.49\textwidth}
   \includegraphics[width=\linewidth]{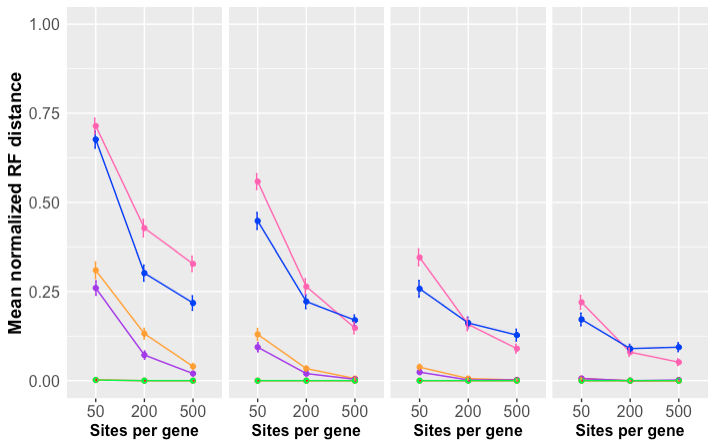}
   \caption{$MBL=0.5$}
\end{subfigure}
\newline
   \begin{subfigure}[b]{0.49\textwidth}
   \includegraphics[width=\linewidth]{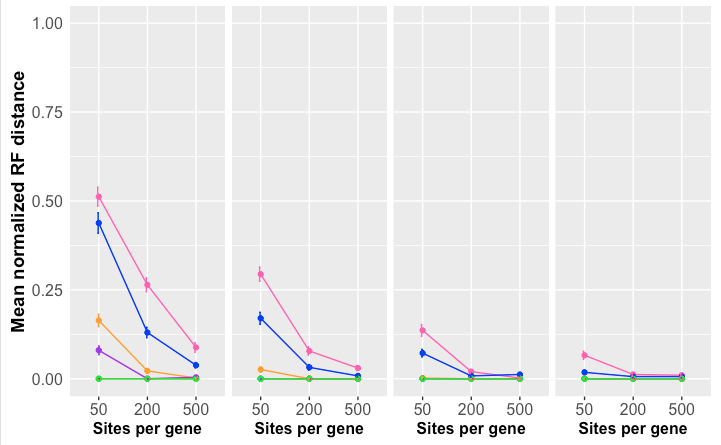}
   \caption{$MBL=1.0$}
\end{subfigure}
\begin{subfigure}[b]{0.49\textwidth}
   \includegraphics[width=\linewidth]{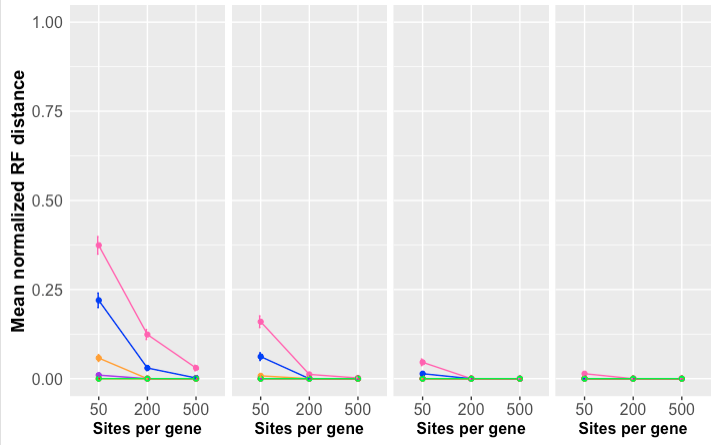}
   \caption{$MBL=2.0$}
\end{subfigure}
\caption{Mean normalized RF distance for Lily-Q vs. ASTRAL for true tree $(((a,b),c),((d,e),(f,(g,h))))$ at different minimum branch lengths (MBL). In each panel, $\theta=\{0.0008, 0.0024, 0.0072, 0.0216\}$ from left to right. Number of genes: 10 -- magenta/blue, 50 -- orange/purple, 500 -- red/green. ASTRAL is displayed in ``hot" colors (magenta/orange/red) and Lily-Q in ``cold" (blue/purple/green). }
\label{simulation_RF6}
\end{figure}

\begin{figure}
\centering
   \begin{subfigure}[b]{0.49\textwidth}
   \includegraphics[width=\linewidth]{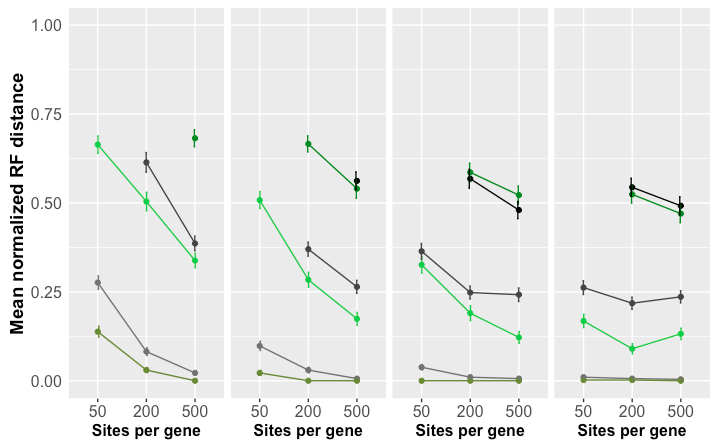}
   \caption{$MBL=0.2$}
\end{subfigure}
\begin{subfigure}[b]{0.49\textwidth}
   \includegraphics[width=\linewidth]{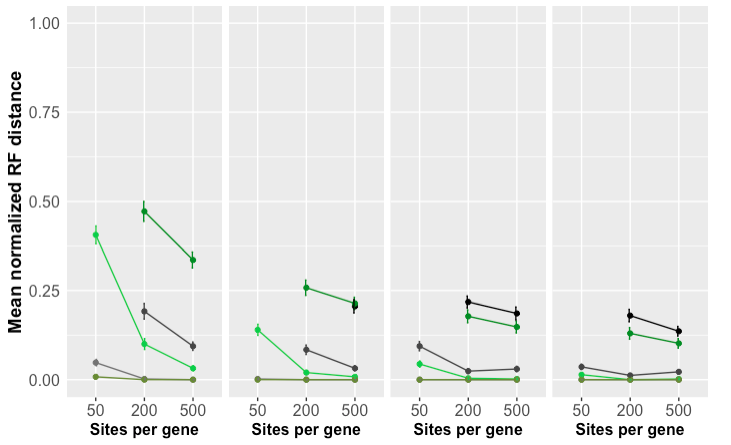}
   \caption{$MBL=0.5$}
\end{subfigure}
\newline
   \begin{subfigure}[b]{0.49\textwidth}
   \includegraphics[width=\linewidth]{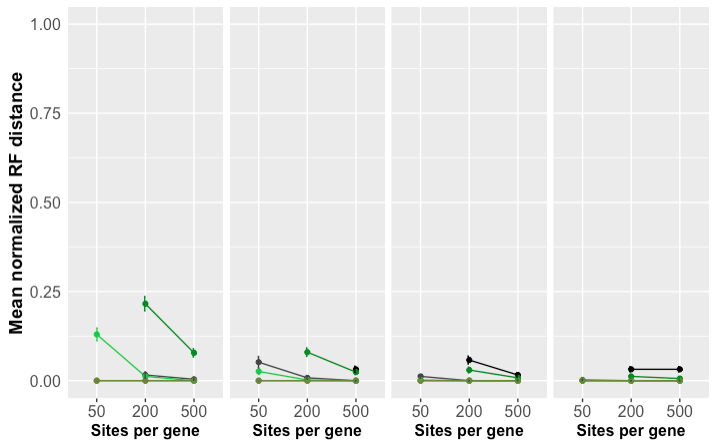}
   \caption{$MBL=1.0$}
\end{subfigure}
\begin{subfigure}[b]{0.49\textwidth}
   \includegraphics[width=\linewidth]{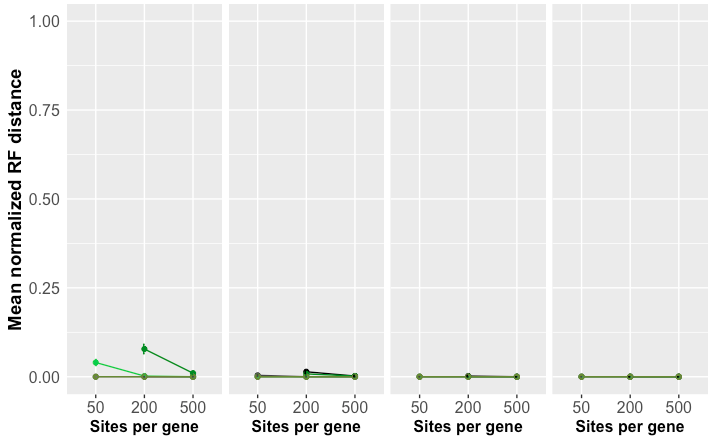}
   \caption{$MBL=2.0$}
\end{subfigure}
\caption{Mean normalized RF distance for Lily-T vs. SVDQuartets for true tree $(((a,b),c),((d,e),(f,(g,h))))$ at different minimum branch lengths (MBL). In each panel, $\theta=\{0.0008, 0.0024, 0.0072, 0.0216\}$ from left to right. Number of genes: 10 -- dark, 50 -- medium, 500 -- light. SVDQuartets displayed in grey shading and Lily-T in green.}
\label{simulation_RF6s}
\end{figure}

\begin{figure}
\centering
   \begin{subfigure}[b]{0.49\textwidth}
   \includegraphics[width=\linewidth]{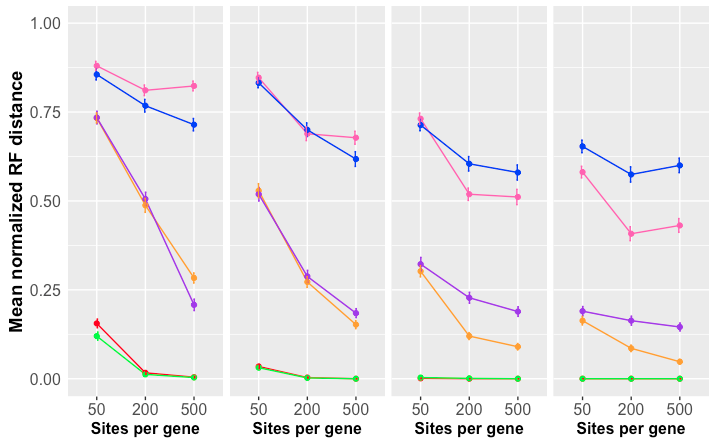}
   \caption{$MBL=0.2$}
\end{subfigure}
\begin{subfigure}[b]{0.49\textwidth}
   \includegraphics[width=\linewidth]{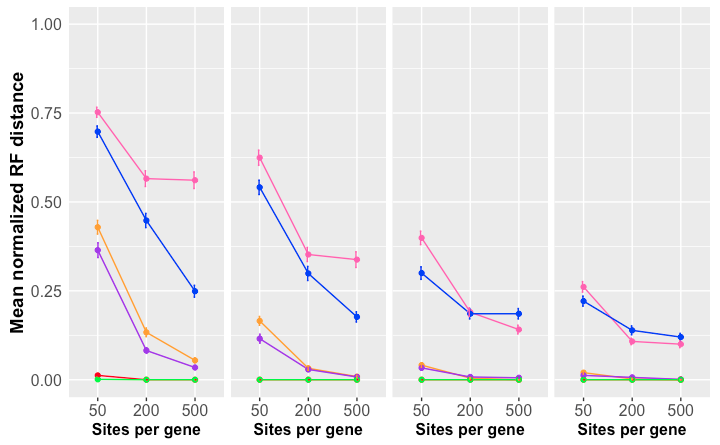}
   \caption{$MBL=0.5$}
\end{subfigure}
\newline
   \begin{subfigure}[b]{0.49\textwidth}
   \includegraphics[width=\linewidth]{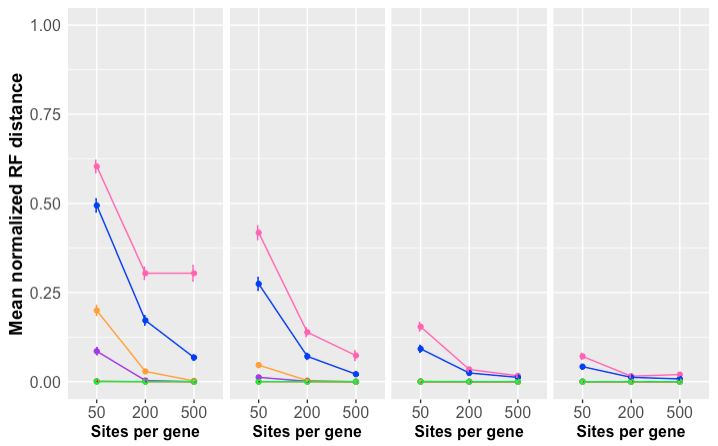}
   \caption{$MBL=1.0$}
\end{subfigure}
\begin{subfigure}[b]{0.49\textwidth}
   \includegraphics[width=\linewidth]{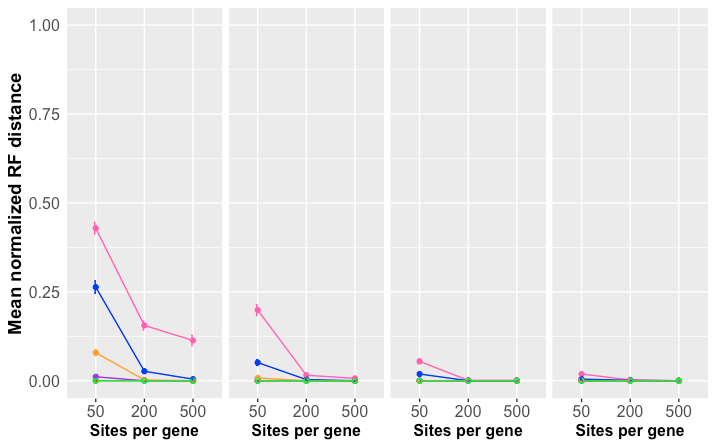}
   \caption{$MBL=2.0$}
\end{subfigure}
\caption{Mean normalized RF distance for Lily-Q vs. ASTRAL for true tree $(((a,(b,c)),(d,(e,f))),((g,(h,i)),(j,(k,l))))$ at different minimum branch lengths (MBL). In each panel, $\theta=\{0.0008, 0.0024, 0.0072, 0.0216\}$ from left to right. Number of genes: 10 -- magenta/blue, 50 -- orange/purple, 500 -- red/green. ASTRAL is displayed in ``hot" colors (magenta/orange/red) and Lily-Q in ``cold" (blue/purple/green). }
\label{simulation_RF12}
\end{figure}

A representative selection of the simulation results are shown in figures \ref{simulation_RF1} through \ref{simulation_RF12} (the full plots are in the supplementary material.) One complication is that for very small number of sites and a small value of $\theta$, there may be cases where there is no divergence between any set of three species. The proper inference in this case would be to return a polytomous tree. Each method, however, treats this case differently. SVDQuartets will either return a polytomous tree or an error. Lily-Q infers an 1/3 probability for each of the three possible unrooted quartets involving three or more zero-divergence species, but the assembly procedure will then produce an error message. Lily-T will do the same, but will generate an error only after consuming much time and memory. ASTRAL will infer an apparently random binary tree, but that is due to FastTree incorrectly inferring binary trees for the gene tree inputs. As a result, for those settings where this occurred, we display the Lily-Q vs. ASTRAL comparison for first 100 valid runs without any set of three zero-distance taxa. We also did not use Lily-T or SVDQuartets on the 10 genes and 50 sites-per-gene setting because SVDQuartets invariably produced errors and Lily-T produced errors while materially slowing down the simulation process. Thus, the plots for Lily-T and SVDQuartets do not show results for some settings. 

A number of results become evident from the RF plots. First, all estimation is better with a larger $\theta$, as there is more mutation along each branch allowing us to pick up more of the phylogenetic signal. A lone exception to this trend is that estimation gets worse for SVDQuartets for $S_8$ with $\theta=0.0216$ and a minimum branch length of 2.0 coalescent units. It is worth noting that for these settings the root is 22.0 coalescent units in the past and the sequences may be nearing saturation and so SVDQuartets may have greater difficulty resolving the phylogeny near saturation. Second, with enough data, all methods perform well. Third, by comparing the 50 genes/500 sites per gene and the 500 genes/50 sites per gene case, we can see that all methods do better with more genes even when the total number of sites is held constant. At smaller sample sizes, SVDQuartets is generally outperformed by all other methods, and Lily-T is in turn outperformed by Lily-Q under most simulation settings.

Therefore, we focused on the comparison between Lily-Q and ASTRAL. For MBL of 1.0 or 2.0 coalescent units, Lily-Q generally performs no worse, and in many cases, better than ASTRAL. For an MBL of 0.5 coalescent units, ASTRAL does better for 200 or 500 sites per gene and $\theta$ of 0.0072 or 0.0216, and this effect appears to be stronger with fewer taxa. Lily-Q does as well or better for almost all cases with 50 sites per gene and 10 or 50 genes. For 0.2 CI, Lily-Q does better with fewer sites per gene and with smaller values of $\theta$ and ASTRAL performs better with more sites per gene and larger values of $\theta$. The comparison also seems somewhat dependent on topology, but without any clear pattern. ASTRAL outperforms Lily-T for most settings, except when the branch lengths are large and $\theta$ is smaller. 

These results match our expectations, as ASTRAL depends on accurate gene tree inputs and those gene trees are easier to resolve when the genes are long (many sites per gene) and there is more mutation along each branch (higher $\theta$). One oddity is that for Lily-Q, with 10 genes, sometimes the RF distance was higher going from 200 to 500 sites per gene. But, the standard errors almost always overlapped so this appears to be random noise arising out of the fact that, with only 10 genes, if the gene trees differ from the species tree, increasing the number of sites only allow better inference of the mismatched gene trees rather than the underlying species tree.

\subsection{Bootstrapping results}
\begin{figure}
    \centering
   \begin{subfigure}{0.49\linewidth} \centering
     \includegraphics[scale=0.3]{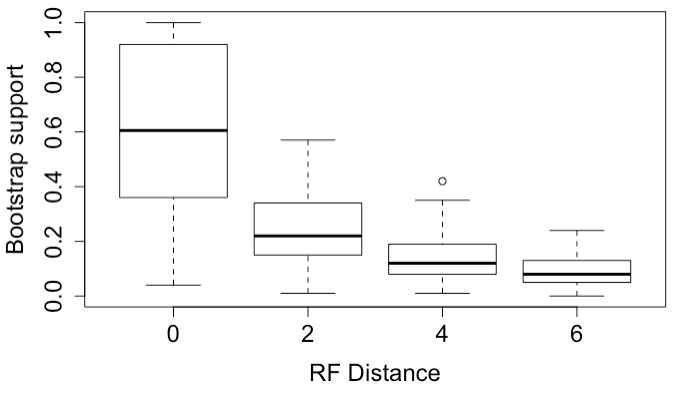}
     \caption{Lily-T (5 taxa)}
   \end{subfigure}
      \begin{subfigure}{0.49\linewidth} \centering
     \includegraphics[scale=0.3]{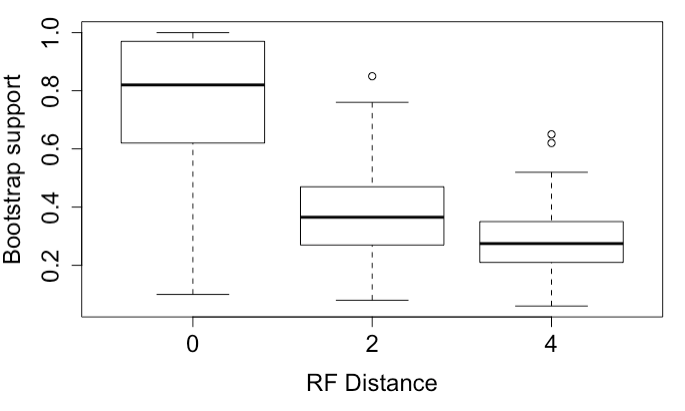}
     \caption{Lily-Q (5 taxa)}
   \end{subfigure}
   \newline
      \begin{subfigure}{0.49\linewidth} \centering
     \includegraphics[scale=0.3]{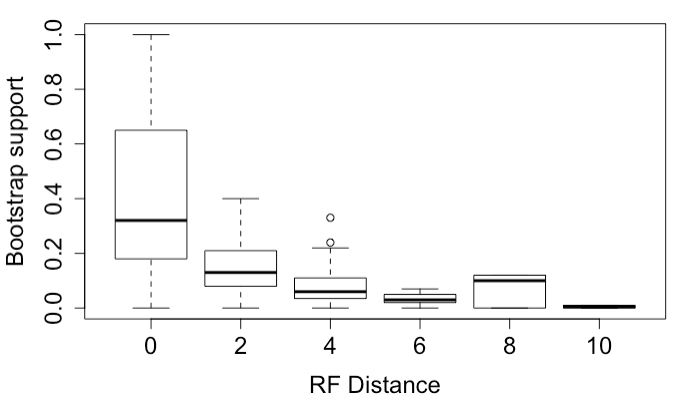}
     \caption{Lily-T (8 taxa)}
   \end{subfigure}
      \begin{subfigure}{0.49\linewidth} \centering
     \includegraphics[scale=0.3]{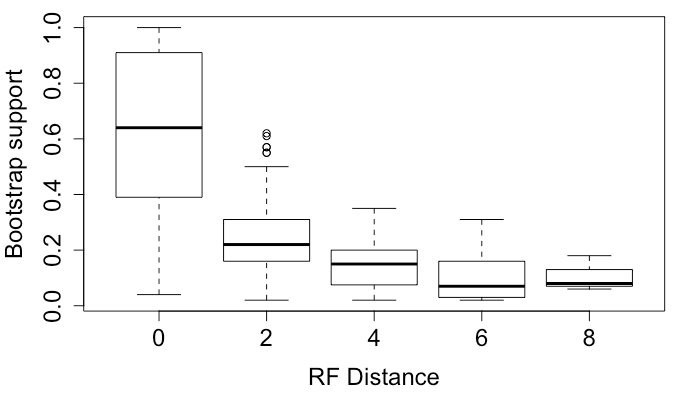}
     \caption{Lily-Q (8 taxa)}
   \end{subfigure}
   \newline
      \begin{subfigure}{0.49\linewidth} \centering
     \includegraphics[scale=0.3]{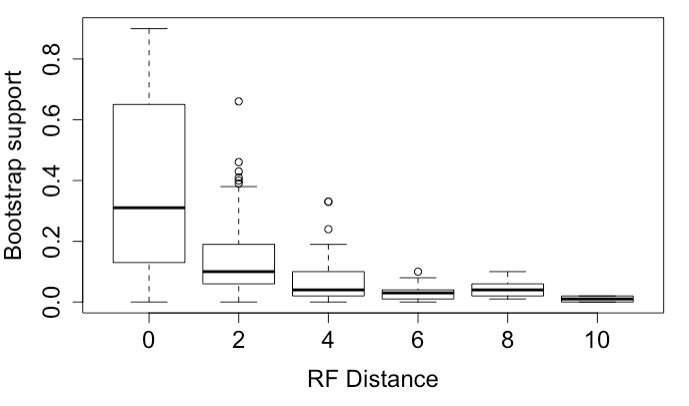}
     \caption{Lily-T (12 taxa)}
   \end{subfigure}
      \begin{subfigure}{0.49\linewidth} \centering
     \includegraphics[scale=0.3]{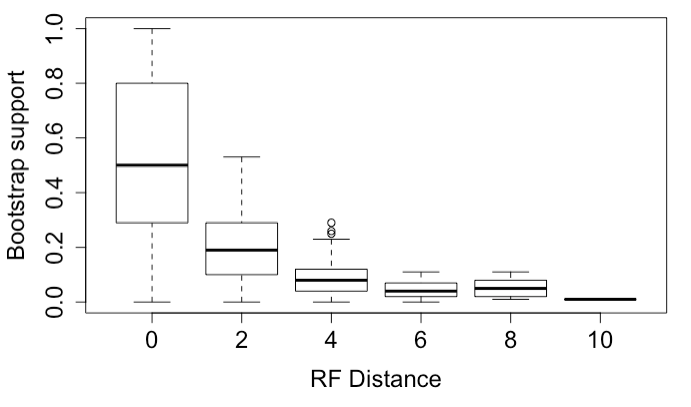}
     \caption{Lily-Q (12 taxa)}
   \end{subfigure}
    \caption{Bootstrap support for estimated trees grouped by RF distance}
    \label{RFboot5}
\end{figure}

For twenty different combinations of topology, $\theta$, MBL, gene length, and number of genes, we drew 100 bootstrap samples by resampling both the genes and then resampling the sites within each gene. For both Lily-T and Lily-Q, we compared the bootstrap support for the estimated tree to the RF distance from the true tree. Figure \ref{RFboot5} shows the bootstrap support given the RF distance between the estimated tree and the true tree. We can see that the farther the estimated tree is from the true tree, the lower the bootstrap support tends to be. Tables \ref{boot_table5} to \ref{boot_table12Q} in turn show the positive and negative predictive value of high or low bootstrap support. What we see is that we have very good positive predictive value -- for Lily-Q only 5 of 697 trees ($<$1\%) with bootstrap support greater than 0.7 were incorrect, and all of them had an RF distance of 2. With lower bootstrap support, however, more trees tend to be further from the true tree but the estimated tree may still be close to the correct tree -- for Lily-Q 356 of 830 estimated trees (43\%) with bootstrap support less than 0.4 were in fact correct.

These simulation studies suggest that the bootstrap support may be useful as a guide to triaging larger phylogenetic questions -- if the support is high, the faster results that can be obtained from Lily-T or Lily-Q can be accepted with confidence. Conversely, if the support is low, it serves as an indication that a more computationally intensive MCMC-based method may be necessary for accurate results.  

\begin{table}[]
    \centering
    \begin{minipage}{0.59\linewidth}
        \begin{tabular}{|c|c|c|c|c|}
        \hline
        \textbf{Bootstrap support} & \textbf{RF 0} & \textbf{RF 2} & \textbf{RF 4} & \textbf{RF 6} \\
         \hline
       0  & 0  & 0  & 0  & 1\\
        \hline
       (0, 0.4]  & 121  & 135  & 109  & 34\\
        \hline
       (0.4, 0.7]  & 107  & 22  & 1  & 0\\
        \hline
       (0.7, 0.9]  & 62  & 0  & 0  & 0\\
        \hline
       (0.9, 1]  & 108  & 0  & 0  & 0 \\
        \hline
    \end{tabular}
    \end{minipage}
    \begin{minipage}{0.39\linewidth}
        \begin{tabular}{|c|c|c|c|}
         \hline
        \textbf{Bootstrap support} & \textbf{RF 0} & \textbf{RF 2} & \textbf{RF 4}  \\
         \hline
       0  & 0  & 0  & 0  \\
        \hline
       (0, 0.4]  & 46  & 88  & 62  \\
        \hline
       (0.4, 0.7]  & 114  & 49  & 12  \\
        \hline
       (0.7, 0.9]  & 153  & 5  & 0  \\
        \hline
       (0.9, 1]  & 171  & 0  & 0  \\
        \hline
    \end{tabular}
    \end{minipage}
    \caption{RF distances of estimated 5-taxon trees by bootstrap support level for Lily-T (left) and Lily-Q (right).}
    \label{boot_table5}
\end{table}

\begin{table}[]
    \centering
        \begin{tabular}{|c|c|c|c|c|c|c|}
         \hline
        \textbf{Bootstrap support} & \textbf{RF 0} & \textbf{RF 2} & \textbf{RF 4} & \textbf{RF 6} & \textbf{RF 8} & \textbf{RF 10}  \\
         \hline
       0  & 1  & 1  & 4  & 1 & 1 & 1 \\
        \hline
       (0, 0.2]  & 129  & 117  & 72 & 19 & 3 & 1  \\
        \hline
       (0.2, 0.9] & 205  & 41  & 3 & 0 & 0 & 0  \\
        \hline
       (0.9, 1] & 100  & 0  & 0 & 0 & 0 & 0   \\
        \hline
    \end{tabular}
    \caption{RF distances of 8-taxon trees estimated by Lily-T for a given bootstrap support level}
    \label{boot_table8T}
\end{table}

\begin{table}[]
    \centering
        \begin{tabular}{|c|c|c|c|c|c|}
         \hline
        \textbf{Bootstrap support} & \textbf{RF 0} & \textbf{RF 2} & \textbf{RF 4} & \textbf{RF 6} & \textbf{RF 8}  \\
         \hline
       (0, 0.4]  & 143  & 113  & 47 & 14 & 3  \\
        \hline
       (0.4, 0.7]  & 149  & 13  & 0 & 0 & 0   \\
        \hline
       (0.7, 0.9] & 87  & 0  & 0 & 0 & 0   \\
        \hline
       (0.9, 1] & 131  & 0  & 0 & 0 & 0    \\
        \hline
    \end{tabular}
    \caption{RF distances of 8-taxon trees estimated by Lily-Q for a given bootstrap support level}
    \label{boot_table8Q}
\end{table}

\begin{table}[]
    \centering
        \begin{tabular}{|c|c|c|c|c|c|c|}
         \hline
        \textbf{Bootstrap support} & \textbf{RF 0} & \textbf{RF 2} & \textbf{RF 4} & \textbf{RF 6} & \textbf{RF 8} & \textbf{RF 10}  \\
         \hline
       0  & 3  & 5  & 3  & 2 & 0 & 1 \\
        \hline
        (0, 0.2]  & 131  & 108  & 55  & 27 & 6 & 1 \\
        \hline
       (0.2, 0.4]  & 68  & 25  & 3 & 0 & 0 & 0  \\
        \hline
       (0.4, 0.7] & 95  & 4  & 0 & 0 & 0 & 0  \\
        \hline
       (0.7, 1] & 63  & 0  & 0 & 0 & 0 & 0   \\
        \hline
    \end{tabular}
    \caption{RF distances of 12-taxon trees estimated by Lily-T for a given bootstrap support level}
    \label{boot_table12T}
\end{table}

\begin{table}[]
    \centering
        \begin{tabular}{|c|c|c|c|c|c|c|}
         \hline
        \textbf{Bootstrap support} & \textbf{RF 0} & \textbf{RF 2} & \textbf{RF 4} & \textbf{RF 6} & \textbf{RF 8} & \textbf{RF 10}  \\
         \hline
       0  & 1  & 2  & 1  & 1 & 0 & 0 \\
        \hline
        (0, 0.2]  & 75  & 52  & 36  & 8 & 6 & 1 \\
        \hline
       (0.2, 0.4]  & 91  & 36  & 4 & 0 & 0 & 0  \\
        \hline
       (0.4, 0.7] & 129  & 7  & 0 & 0 & 0 & 0  \\
        \hline
      (0.7, 0.9] & 88  & 0  & 0 & 0 & 0 & 0  \\
        \hline
       (0.9, 1] & 62  & 0  & 0 & 0 & 0 & 0   \\
        \hline
    \end{tabular}
    \caption{RF distances of 12-taxon trees estimated by Lily-Q for a given bootstrap support level}
    \label{boot_table12Q}
\end{table}

\subsection{Application to empirical sequence data}
\begin{figure}
    \centering
   \begin{subfigure}{0.49\linewidth} \centering
     \includegraphics[scale=0.4]{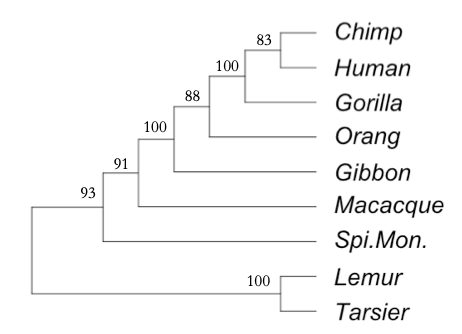}
     \caption{Lily-T}
   \end{subfigure}
      \begin{subfigure}{0.49\linewidth} \centering
     \includegraphics[scale=0.4]{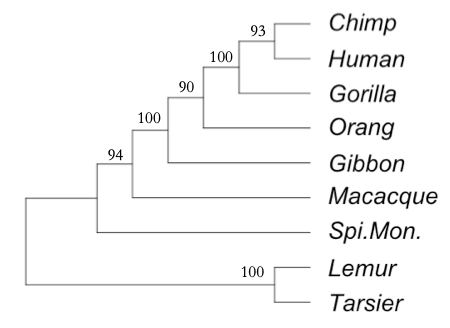}
     \caption{Lily-Q}
   \end{subfigure}
    \caption{Estimates for 9-taxon primate tree with bootstrap support for each clade. Lily-Q display assumes root location is known.}
    \label{primate_out}
\end{figure}

We applied Lily-T and Lily-Q to four empirical datasets. For all the datasets we estimated the topology and then drew 100 bootstrap samples and calculated the bootstrap support at each node. The first dataset is a somewhat toy example: a segment of 888 sites over five genes coming from the mitochondria of nine primates \cite{yang97}. Since mitochondria are passed down solely from the mother, we performed the bootstrap as if the data were from a single 888-site gene since all sites share a common lineage. The results are displayed in figure \ref{primate_out}. Both Lily-T and Lily-Q inferred the same tree as \cite{yang97}. This tree is different than the primate consensus from \cite{chou15, gatesy17} in that infers that tarsiers and lemurs are a cherry rather than that lemurs are an outgroup, likely due to the small number of sites from a single lineage. Lily-Q had higher bootstrap support than Lily-T for most nodes, especially for the (human, chimp) clade. Run time for both methods was a matter of seconds, including bootstrapping.

\begin{figure}
    \centering
   \begin{subfigure}{0.49\linewidth} \centering
     \includegraphics[scale=0.4]{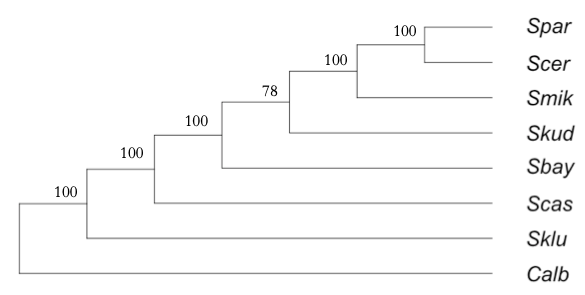}
     \caption{Lily-T}
   \end{subfigure}
      \begin{subfigure}{0.49\linewidth} \centering
     \includegraphics[scale=0.4]{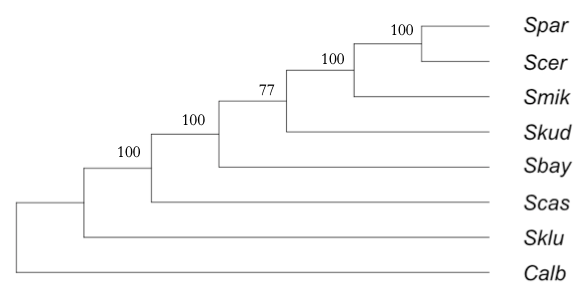}
     \caption{Lily-Q}
   \end{subfigure}
    \caption{Estimates for 8-taxon yeast tree with bootstrap support for each clade. Lily-Q display assumes root location is known.}
    \label{yeast_out}
\end{figure}

The second dataset consisted of approximately 127,000 sites over 106 genes from eight yeast species: \textit{S.cerevisiae} (Scer), \textit{S. paradoxus} (Spar), \textit{S. mikatae} (Smik), \textit{S. kudriavzevii} (Skud), \textit{S. bayanus} (Sbay), \textit{S. castellii} (Scas), \textit{S. kluyveri} (Sklu), and the outgroup \textit{C. albicans} (Calb) \cite{rokas03, wen18}. Both methods matched the results from \cite{wen18}. Estimation took approximately 4 min including bootstrapping on a MacBook Air. All but one node had 100\% bootstrap support. The exception was some uncertainty over whether \textit{Skud} and \textit{Sbay} represent a clade or if \textit{Sbay} is an outgroup to all of \textit{Scer}, \textit{Spar}, and \textit{Smik} (see figure \ref{yeast_out}).

\begin{figure}
    \centering
   \begin{subfigure}{0.49\linewidth} \centering
     \includegraphics[scale=0.4]{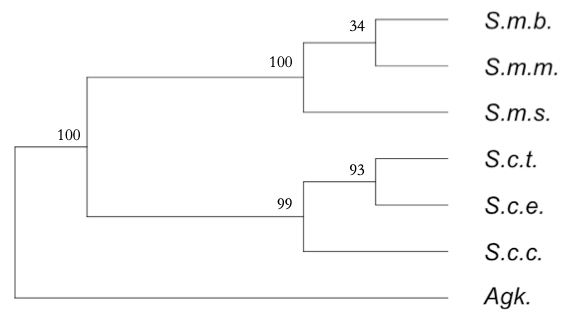}
     \caption{Lily-T}
   \end{subfigure}
      \begin{subfigure}{0.49\linewidth} \centering
     \includegraphics[scale=0.4]{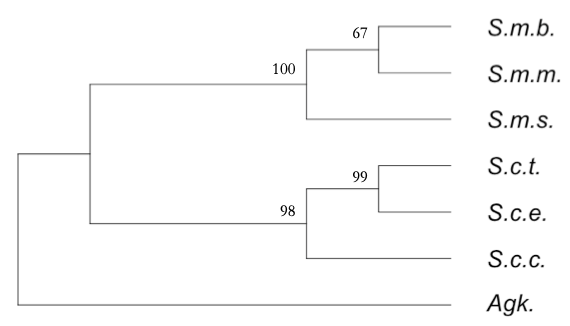}
     \caption{Lily-Q}
   \end{subfigure}
    \caption{Estimates for 7-taxon snake tree with bootstrap support for each clade. (Species names abbreviated for clarity. Full names in \cite{chifman14}). Lily-Q display assumes root location is known.}
    \label{snake_out}
\end{figure}

Next, we applied Lily-T and Lily-Q to a dataset of 52 individuals from seven North American snake species consisting of aligned sequences from 18 nuclear and 1 mitochondrial genes with 190-850 characters per gene. Runtime was around a half hour on a Unix HPC platform (the increase in run time was due to the many different combination of individuals within each subset of three or four species). This is much less than the runtimes reported by \cite{chifman14}: $<$1 day for SVDQuartets and $\approx$10 days for *BEAST, although we acknowledge runtimes for these methods have improved in the time since those results were published. Our results are shown in figure \ref{snake_out}. Bootstrap support was over 90\% for all clades except the (S.m.miliarius, S.m.barbouri) clade. For that clade, bootstrap support was 34\% for Lily-T and 67\% for Lily-Q. SVDQuartets also exhibited low bootstrap support for this clade, so the weak support may be a function of the weak phylogenetic signal present in the data.

\begin{figure}
    \centering
   \begin{subfigure}{0.49\linewidth} \centering
     \includegraphics[scale=0.3]{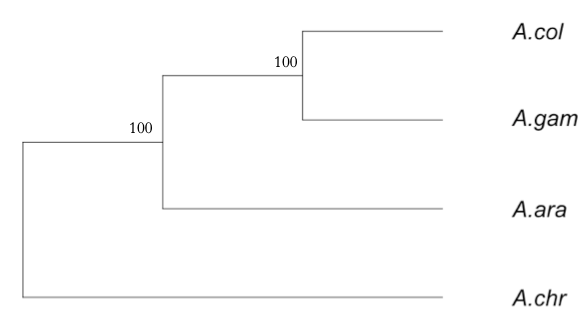}
     \caption{Lily-T}
   \end{subfigure}
      \begin{subfigure}{0.49\linewidth} \centering
     \includegraphics[scale=0.3]{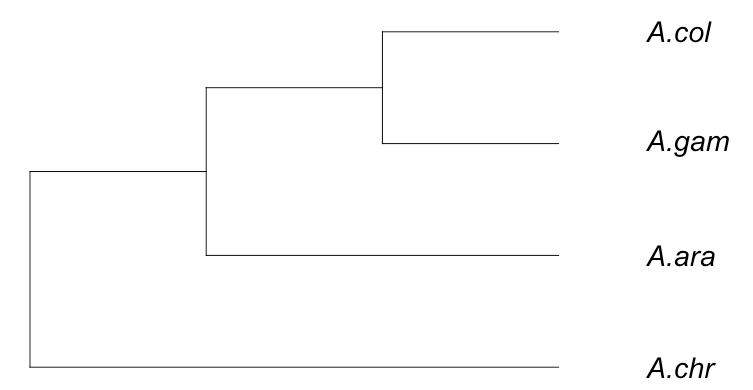}
     \caption{Lily-Q}
   \end{subfigure}
    \caption{Estimates for 4-taxon yeast mosquito with bootstrap support (Lily-T). Lily-Q display assumes root location is known (posterior probability of 100\% not shown).}
    \label{mos_out}
\end{figure}

Finally, we estimated the topology for four mosquito species using aligned sequences consisting of over 25 million sites from around 80,000 different genes. For Lily-T, we ran 100 bootstrap samples and there was 100\% bootstrap support for each node, with total computation time under an hour using a Unix HPC platform even with this large dataset. Because we did not have the delineation of the different genes, we resampled the sites for bootstrapping in a single stage. The estimated species tree is shown in figure \ref{mos_out} and matched the results from \cite{tha18}. Since there were only four taxa, Lily-Q could calculate a posterior probability of 100\% using a single run which took around 15 seconds. 

\section{Conclusions and further work}
Lily-T and Lily-Q are two new methods for fast, accurate species tree estimation under the multispecies coalescent model along with the assumption of the molecular clock. The methods are insensitive to the value of the coalescent population parameter $\theta$, and sensitivity to the prior on branching times can be controlled through our method for estimating root age from the data. Lily-Q is more accurate than Lily-T, but Lily-T does have certain advantages in that it can estimate the root location, and there is a theoretical guarantee as the number of unlinked sites goes to infinity.

The comparison between Lily-Q and ASTRAL follows our expectations in that ASTRAL is able to perform well under conditions when estimation of the individual genes is likely to be accurate -- when $\theta$ is large and when there are a large number of sites per gene. It is worth noting that our comparisons are of a correctly specified ASTRAL model against an incorrectly specified Lily-Q model: we assume that we have delineated the genes correctly and that sites within each gene have a common history whereas Lily-Q assumes that all sites are unlinked. Even with this disadvantage, Lily-Q outperformed ASTRAL for many parameter settings. There is also reason to believe that the number of sites per gene needed for ASTRAL to do better than Lily-Q may not be realistic -- \cite{hobolth11} estimated 75\% of loci in a primate dataset have a recombination-free length of between 17 and 93bp and \cite{springer16} reported a loci length of around 12bp for a large mammalian dataset. An avenue for further research would be to test Lily-Q against ASTRAL when the genes are not properly delimited to measure how much this degrades ASTRAL performance.

We also did not compare directly to any concatenation methods. These comparisons may indicate the limits of concatenation methods, especially for the highly asymmetric trees with short MBL. One comparison in particular that may be of interest would be to compare Lily-T to SMRT-ML \cite{SMRTML}, which estimates rooted triplets using concatenated data. We should highlight that, unlike concatentation or coestimation methods, we can only estimate $S$, and not any other parameters of interest such as $\boldsymbol{\tau}$, although other methods for estimating $\boldsymbol{\tau}$ incorporating ILS such as that of \cite{peng20} may be able to be used in conjunction with our work.

\begin{figure}
            \centering
        \includegraphics[width=0.5\linewidth]{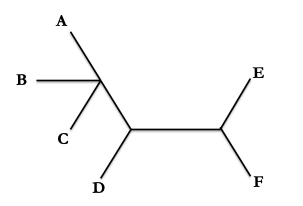}
    \caption{Handling identical sequences: After first treating the identical sequences as coming from a single species, the rest of the tree is resolved as $(c,d,(e,f))$. Then $c$ is replaced by the polytomy $(a,b,c)$.}
    \label{polytomy}
\end{figure}

While both the Lily-T and Lily-Q assembly steps generate errors in the face of three or more identical sequences, there are ways to work around this problem. For example, consider the case with six taxa where taxa $a$, $b$, and $c$ have identical sequences. As a first step $a$ and $b$ could be simply removed from consideration, inference applied to the remaining taxa (perhaps inferring the split $(c,d,(e,f))$, which results in the final inferred tree in figure \ref{polytomy}. While such a workaround would be unwieldy for the large-scale simulation study performed here, it could be reasonably implemented on empirical data.

We have presented some data showing that high bootstrap support is indicative of high tree estimation accuracy, but we wish to highlight that we need to explore this claim further in future work. First, to minimize simulation time, we only performed the bootstrap analysis on twenty different parameter settings. Second, we would like to perform further simulation studies to investigate whether bootstrap support for a particular clade has predictive value for the clade being present in the true tree.

Finally, this work has all been done under the molecular clock assumption, and we have some early signs that the methods are in fact quite sensitive to this assumption. To see why, consider a true tree $((a,b),c)$ with the branch leading to $a$ having a much larger mutation rate than other branches. This would lead to a larger number of YXX patterns than under equal rates, which if the divergence in rates was large enough, would lead Lily-T to infer $(a,(b,c))$ rather than the true tree. We hope to investigate this in greater detail in future work. Correcting for sensitivity to violations of the clock would require deriving $\boldsymbol{\delta}|(S,\boldsymbol{\tau},\boldsymbol{\gamma}) $ where $\boldsymbol{\gamma}$ is the vector of relative rates along different branches of the tree. Then we would need to either estimate $\hat{\boldsymbol{\gamma}}$ from the data or apply a prior to $\boldsymbol{\gamma}$ and integrate over it to get $\boldsymbol{\delta}|(S,\boldsymbol{\tau}) $. Whether these steps can be performed without substantially slowing down the methodology remains an open question.

\section{Appendix}

\subsection{Calculation of $\boldsymbol{\delta}|(S,\boldsymbol{\tau}))$ for 3-taxon trees}
\label{tripletApp}
The fifteen site pattern probabilities derived in \cite{chifman15} can be mapped down to the five site patterns for three taxa by summing over the fourth taxon we are not interested in to get the following relationships (dropping the conditioning on $(S, \boldsymbol{\tau})$ for clarity):
$$p_0=\delta_{XXX}=\delta_{XXXX}+3\delta_{YXXX}$$
$$p_1=\delta_{YXX}=\delta_{XYXX}+\delta_{XXYY}+2\delta_{YZXX}$$
$$p_2=\delta_{XXY}=\delta_{XYX}=\delta_{XXXY}+\delta_{XYYX}+2\delta_{YXXZ}=\delta_{XXYX}+\delta_{XYXY}+2\delta_{YXZX}$$
$$p_3=\delta_{XYZ}=\delta_{XXYZ}+\delta_{XYXZ}+\delta_{XYZX}+\delta_{XYZW}$$
These probabilities take the form:
\begin{eqnarray}
\label{3taxaProb1}
  \left\{
  \begin{aligned}
& p_0=c_0+3c_1+6c_2+12c_3\\
& p_1=c_0+3c_1-2c_2-4c_3\\
& p_2=c_0-c_1+2c_2-4c_3\\
& p_3=c_0-c_1-2c_2+4c_3
  \end{aligned}
  \right.
\end{eqnarray}
where (measuring $\tau_1$ and $\tau_2$ in coalescent units):
\begin{eqnarray}
\label{3taxaProb2}
  \left\{
  \begin{aligned}
& c_0=1/64\\
& c_1=\frac{e^{-8\tau_1\theta/3}}{64(1+\frac{8\theta}{3})}\\
& c_2=\frac{e^{-8\tau_2\theta/3}}{64(1+\frac{8\theta}{3})}\\
& c_3=\frac{e^{-4\tau_1\theta/3}e^{-8\tau_2\theta/3}}{128(1+\frac{8\theta}{3})(1+\frac{4\theta}{3})}
  \end{aligned}
  \right.
\end{eqnarray}
It is easily verified that:
\begin{equation}
\label{3taxaProb3}
    4p_0+12p_1+24p_2+24p_3=1
\end{equation}
The coefficients in equation \ref{3taxaProb3} arise as follows. $X$ can represent any of $A$, $C$, $G$, or $T$. $Y$ can represent any of the remaining three characters, and $Z$ any of the remaining two. Finally, the coefficient of $p_2$ is doubled to account for both $\delta_{XYX}$ and $\delta_{XXY}$.

\subsection{Calculation of $\boldsymbol{\delta}|S$ for 3-taxon trees}
One can see from equation \ref{3taxaProb1} that $\delta_k|(S,\boldsymbol{\tau})$ is a linear combination of terms, so $\delta_k|S$ is also linear combination of terms where each integral is of one of two forms (since the prior on $\boldsymbol{\tau}$ is $f(\boldsymbol{\tau}) = \frac{1}{\tau_2}\beta e^{-\beta\tau_2}$): 

\textbf{Form one:}
\begin{equation}
\label{post1}
    \int_0^{\infty} \int_0^{\tau_2} \frac{c}{\tau_2}e^{-a\tau_2} d\tau_1 d\tau_2=\int_0^{\infty} ce^{-a\tau_2}d\tau_2=\frac{c}{a}
\end{equation}

\textbf{Form two:}
$$\int_0^{\infty} \int_0^{\tau_2} \frac{c}{\tau_2}e^{-b\tau_1}e^{-a\tau_2} d\tau_1 d\tau_2$$
$$=\int_0^{\infty}\frac{c}{b\tau_2} (1-e^{-b\tau_2}) e^{-a\tau_2}d\tau_2$$
\begin{equation}
\label{post2}
    =\frac{c}{b}log(\frac{a+b}{a})
\end{equation}
Taken together, $\delta_k|S$ has the linear form of equation \ref{3taxaProb1} with the following terms replacing those of equation \ref{3taxaProb2}: 
\begin{eqnarray}
\label{3taxaProb4}
  \left\{
  \begin{aligned}
& c_0=1/64\\
& c_1=\frac{\beta}{64(1+\frac{8\theta}{3})(\frac{8\theta}{3})}log(\frac{\beta+\frac{8\theta}{3}}{\beta})\\
& c_2=\frac{\beta}{64(1+\frac{8\theta}{3})(\beta+\frac{8\theta}{3})}\\
& c_3=\frac{\beta}{64(1+\frac{8\theta}{3})(2+\frac{8\theta}{3})(\frac{4\theta}{3})}log(\frac{\beta+4\theta}{\beta+\frac{8\theta}{3}})
  \end{aligned}
  \right.
\end{eqnarray}
These results were verified by Monte Carlo integration with 10,000 draws from the prior on $\boldsymbol{\tau}$ with $\beta=0.1$.

\subsection{Calculation of $\boldsymbol{\delta}|S$ for 4-taxon trees}

From Chifman and Kubatko (2015), $P(\boldsymbol{\delta}|(S,\boldsymbol{\tau}))$ is the inner product $\textbf{c}^{T}\textbf{b}$ where $\textbf{c}$ is a vector of constants (with respect to the branching times) and $\textbf{b}$ is a vector of functions of the branching times. These two vectors (where $\alpha=4/3$ is a constant that comes from the mutation rate when time is measured in coalescent units) are:
$$\boldsymbol{b}_{sym}=\{1,e^{-2\alpha\theta\tau_1},e^{-2\alpha\theta\tau_2}, e^{-2\alpha\theta(\tau_1+\tau_2)}, e^{-2\alpha\theta\tau_3},
           e^{-\alpha\theta(\tau_1+2\tau3)},  e^{-\alpha\theta(\tau_2+2\tau3)},$$  $$e^{-\alpha\theta(\tau_1+\tau_2+2\tau3)},
            e^{2\alpha(\tau_1+\tau_2)-4\theta\tau_3(\alpha+\frac{1}{2\theta})}
            \} $$
$$\boldsymbol{b}_{asymm}=\{1,e^{-2\alpha\theta\tau_1},e^{-2\alpha\theta\tau_2}, e^{-\alpha\theta(\tau_1+2\tau_2)}, e^{-2\alpha\theta\tau_3},
           e^{-\alpha\theta(\tau_1+2\tau3)}, e^{-2\alpha\theta(\tau_1+\tau3)},$$ $$ e^{-\alpha\theta(\tau_2+2\tau3)},  e^{-\alpha\theta(\tau_1+\tau_2+2\tau3)},
            e^{2(\tau_1-\tau_2)-2\theta\alpha(\tau_2+\tau3)}
            \} $$
Since the priors are $f(\boldsymbol{\tau}) = \frac{1}{\tau_3^2}\beta e^{-\beta\tau_3}$ for the symmetric topology and $f(\boldsymbol{\tau}) = \frac{2}{\tau_3^2}\beta e^{-\beta\tau_3}$ for the asymmetric topology, the integrals required to integrate out the branching times take on only eight different forms (four for each topology).
\subsubsection{Symmetric topology}
Form One:
$$\int_0^{\infty} \int_0^{\tau_3} \int_0^{\tau_3} \frac{c}{\tau_3^2} e^{-f\tau_3} d\tau_1 d\tau_2 d\tau_3=\int_0^{\infty} c e^{-f\tau_3}d\tau_3= \frac{c}{f}$$
Form two:
$$\int_0^{\infty} \int_0^{\tau_3} \int_0^{\tau_3} \frac{c}{\tau_3^2} e^{-a\tau_1} e^{-f\tau_3} d\tau_1 d\tau_2 d\tau_3=\int_0^{\infty} \frac{c}{a\tau_3}(1-e^{-a\tau_3}) e^{-f\tau_3}d\tau_3= \frac{c}{a}(log(a+f)-log(f))$$
After applying integration by parts.\\
Form three
$$\int_0^{\infty} \int_0^{\tau_3} \int_0^{\tau_3} \frac{c}{\tau_3^2} e^{-b\tau_2} e^{-f\tau_3} d\tau_1 d\tau_2 d\tau_3=\int_0^{\infty} \frac{c}{b\tau_3}(1-e^{-b\tau_3}) e^{-f\tau_3}d\tau_3= \frac{c}{b}(log(b+f)-log(f))$$
Form four:
$$\int_0^{\infty} \int_0^{\tau_3} \int_0^{\tau_3} \frac{c}{\tau_3^2}  e^{-a\tau_1} e^{-b\tau_2} e^{-f\tau_3} d\tau_1 d\tau_2 d\tau_3= \int_0^{\infty} \frac{c}{ab\tau_3^2} (1-e^{-a\tau_3})(1-e^{-b\tau_3})e^{-f\tau_3} d\tau_3$$
Use integration by parts to obtain (the first term vanishes):
$$=\frac{c}{ab}\int_0^{\infty}\frac{1}{\tau_3}(-fe^{-f\tau_3}+(a+f)e^{-(a+f)\tau_3}+(b+f)e^{-(b+f)\tau_3}-(a+b+f)e^{-(a+b+f)\tau_3}) d\tau_3$$
After applying integration by parts a second time (again, the first term vanishes):
$$=\frac{c}{ab}(flog(f)-(a+f)log(a+f)-(b+f)log(b+f)+(a+b+f)log(a+b+f))$$
\subsubsection{Asymmetric topology}
Form One:
$$\int_0^{\infty} \int_0^{\tau_3} \int_0^{\tau_2} \frac{c}{\tau_3^2} e^{-f\tau_3} d\tau_1 d\tau_2 d\tau_3=\int_0^{\infty} \frac{c}{2} e^{-f\tau_3}d\tau_3= \frac{c}{2f}$$
Form Two:
$$\int_0^{\infty} \int_0^{\tau_3} \int_0^{\tau_2} \frac{c}{\tau_3^2} e^{-a\tau_1} e^{-f\tau_3} d\tau_1 d\tau_2 d\tau_3 =\int_0^{\infty} \int_0^{\tau_3} \frac{c}{a\tau_3^2} (1-e^{-a\tau_2})e^{-f\tau_3} d\tau_2 d\tau_3$$
$$=\int_0^{\infty} \frac{c}{a^2\tau_3^2}(a\tau_3e^{-f\tau_3}-e^{-f\tau_3}+e^{-(a+f)\tau_3}) d\tau_3$$
$$= \int_0^{\infty} \frac{c}{a\tau_3}e^{-f\tau_3} d\tau_3 + \int_0^{\infty} \frac{c}{a^2\tau_3^2} (e^{-(a+f)\tau_3}-e^{-f\tau_3})$$
Perform integration by parts once on the second integral to obtain:
$$= \int_0^{\infty} \frac{c}{a\tau_3}e^{-f\tau_3} d\tau_3 + \int_0^{\infty} \frac{c}{a^2\tau_3} (fe^{-f\tau_3}-(a+f)e^{-(a+f)\tau_3}) - \frac{c}{a}$$
Grouping terms:
$$= \frac{c(a+f)}{a^2} \int_0^{\infty} \frac{1}{\tau_3} (e^{-f\tau_3}-e^{-(a+f)\tau_3}) - \frac{c}{a}$$
After performing integration by parts again:
$$=\frac{c(a+f)}{a^2}(log(a+f)-log(f))-\frac{c}{a} $$
Form three:
$$\int_0^{\infty} \int_0^{\tau_3} \int_0^{\tau_2} \frac{c}{\tau_3^2} e^{-b\tau_2} e^{-f\tau_3} d\tau_1 d\tau_2 d\tau_3=\int_0^{\infty} \int_0^{\tau_3} \frac{c\tau_2}{\tau_3^2}e^{-b\tau_2}e^{-f\tau_3} d\tau_2d\tau_3$$
$$=\int_0^{\infty} \frac{c}{b^2\tau_3^2}(e^{-d\tau_3}-e^{-(b+d)\tau_3}-b\tau_3e^{-(b+d)\tau_3}) d\tau_3$$
Using integration by parts gives:
$$=\frac{c}{b}+\int_0^{\infty} \frac{cf}{b^2\tau_3} (e^{-(b+f)\tau_3}-e^{-f\tau_3})$$
After applying integration by parts a second time:
$$=\frac{c}{b}+\frac{cf}{b^2}(log(f)-log(b+f)) $$
Form four: 
$$\int_0^{\infty} \int_0^{\tau_3} \int_0^{\tau_2} \frac{c}{\tau_3^2} e^{-a\tau_1} e^{-b\tau_2} e^{-f\tau_3} d\tau_1 d\tau_2 d\tau_3=\int_0^{\infty} \int_0^{\tau_3} \frac{c}{a\tau_3^2} (1-e^{-a\tau_2}) e^{-b\tau_2} e^{-f\tau_3} d\tau_2 d\tau_3$$
$$=\int_0^{\infty} \frac{c}{ab(a+b)\tau_3^2} (ae^{-f\tau_3}-(a+b)e^{-(b+f)\tau_3}+be^{-(a+b+f)\tau_3}) d\tau_3$$
Applying integration by parts (the first term vanishes) gives:
$$= \int_0^{\infty} \frac{c}{ab(a+b)\tau_3} (-afe^{-f\tau_3}+(a+f)(b+f)e^{(b+f)\tau_3}-b(a+b+f)e^{-a+b+f)\tau_3}) d\tau_3$$
After integrating by parts a second time (again the first term vanishes):
$$=\frac{c}{ab(a+b)}(aflog(f)-(a+b)(b+f)log(b+f)+b(a+b+f)log(a+b+f))$$
\subsubsection{Example}
As an example, we will show the calculation of the second term of $\textbf{c}^T\textbf{d}$ for an asymmetric topology. From above, $b_2=e^{-2\alpha\theta\tau_1}$ and so the integral is:
$$\int_0^{\infty} \int_0^{\tau_3} \int_0^{\tau_2} \frac{2\beta c_2}{\tau_3^2} e^{-2\alpha\theta\tau_1} e^{-\beta\tau_3} d\tau_1 d\tau_2 d\tau_3 $$
This is of form two where $c=2\beta c_2$, $a=2\alpha\theta$, and $f=\beta$.
Then, from above,
$$d_2=\frac{2\beta c_2(2\alpha\theta+\beta)}{(2\alpha\theta)^2}(log(2\alpha\theta+\beta)-log(\beta))-\frac{2\beta c_2}{2\alpha\theta}$$
\subsubsection{Final results}
$$d_{sym,1}=1 $$
$$d_{sym,2}=d_{sym,3}=\beta\frac{log(2\alpha\theta+\beta)-log(\beta)}{2\alpha\theta}$$
$$d_{sym,4}= \beta\frac{\beta log(\beta)-2(2\alpha\theta+\beta)log(2\alpha\theta+\beta)+(4\alpha\theta+\beta)log(4\alpha\theta+\beta)}{4\alpha^2\theta^2}$$
$$d_{sym,5}= \beta\frac{1}{2\alpha\theta+\beta}$$
$$d_{sym,6}= d_{sym,7}=\beta\frac{log(3\alpha\theta+\beta)-log(2\alpha\theta+\beta)}{\alpha\theta}$$
$$d_{sym,8}= \beta\frac{(2\alpha\theta+\beta)log(2\alpha\theta+\beta)
-2(3\alpha\theta+\beta)log(3\alpha\theta+\beta)
+(4\alpha\theta+\beta)log(4\alpha\theta+\beta)}{\alpha^2\theta^2}$$
$$d_{sym,9}=\beta(\beta+4\theta(\alpha+\frac{1}{2\theta}))log((\beta+4\theta(\alpha+\frac{1}{2\theta}))
-2(\beta-1+4\theta(\alpha+\frac{1}{2\theta}))log((\beta-1+4\theta(\alpha+\frac{1}{2\theta}))$$
$$+(\beta-2+4\theta(\alpha+\frac{1}{2\theta}))log((\beta-2+4\theta(\alpha+\frac{1}{2\theta}))$$
$$d_{asymm,1}=1 $$
$$d_{asymm,2}= 2\beta\frac{2\alpha\theta+\beta}{4\alpha^2\theta^2}(log(2\alpha\theta+\beta)-log(\beta))-\frac{2\beta}{2\alpha\theta}$$
$$d_{asymm,3}= 2\beta\frac{\beta}{4\alpha^2\theta^2}(log(\beta)-log(2\alpha\theta+\beta))+\frac{2\beta}{2\alpha\theta}$$
$$d_{asymm,4}= \frac{2\beta}{6\alpha^3\theta^3}(\alpha\theta\beta log(\beta)-(3\alpha\theta)(2\alpha\theta+\beta)log(2\alpha\theta+\beta)+(2\alpha\theta)(3\alpha\theta+\beta)log(3\alpha\theta+\beta))$$
$$d_{asymm,5}=\frac{2\beta}{4\alpha\theta+2\beta}$$
$$d_{asymm,6}=2\beta\frac{3\alpha\theta+\beta}{\alpha^2\theta^2}(log(3\alpha\theta+\beta)-log(2\alpha\theta+\beta))-\frac{2\beta}{\alpha\theta}$$
$$d_{asymm,7}=2\beta\frac{4\alpha\theta+\beta}{4\alpha^2\theta^2}(log(4\alpha\theta+\beta)-log(2\alpha\theta+\beta))-\frac{2\beta}{2\alpha\theta}$$
$$d_{asymm,8}=2\beta\frac{2\alpha\theta+\beta}{\alpha^2\theta^2}(log(2\alpha\theta+\beta)-log(3\alpha\theta+\beta))+\frac{2\beta}{\alpha\theta}$$
$$d_{asymm,9}=\frac{2\beta}{2\alpha^3\theta^3}(\alpha\theta(2\alpha\theta+\beta)log(2\alpha\theta+\beta)-(2\alpha\theta)(3\alpha\theta+\beta)log(3\alpha\theta+\beta)+(\alpha\theta)(4\alpha\theta+\beta)log(4\alpha\theta+\beta))$$
$$d_{asymm,10}=\frac{-2\beta}{(2\alpha\theta)(1+2\alpha\theta)}(-(2\alpha\theta+\beta)log(2\alpha\theta+\beta)-(2\alpha\theta)(1+4\alpha\theta+\beta)log(1+4\alpha\theta+\beta)$$
$$+(1+2\alpha\theta)(4\alpha\theta+\beta)log(4\alpha\theta+\beta)) $$

\subsection{Proof of Theorem 1}
\label{mainTheorem}

\begin{proof}
Without loss of generality, let $S_0=(a,(b,c))$ be the true tree and $\boldsymbol{\tau}_0$ be the true value of $\boldsymbol{\tau}$. We want to show that $\forall \epsilon>0 \, \, \exists J_0 : \forall J>J_0 \, \, P(\hat{S}=S)>1-\epsilon$. First, note that for all values of $(\boldsymbol{\tau}, \theta)$, we have $\delta_{XXY}=\delta_{XYX}$. 

Second, we note using the results of equation \ref{3taxaProb1}
\begin{equation}
    \delta_{YXX}|((a,(b,c)),\boldsymbol{\tau}) > \delta_{XYX}|((a,(b,c)),\boldsymbol{\tau})
    \label{expectation}
\end{equation}
$$\leftrightarrow c_0+3c_1-2c_2-4c_3 > c_0-c_1+2c_2-4c_3$$
$$\leftrightarrow \frac{e^{-2m\tau_1\theta}}{64(1+2m\theta)}=c1>c2=\frac{e^{-2m\tau_2\theta}}{64(1+2m\theta)}$$
$$\leftrightarrow \tau_1<\tau_2$$
which holds w.p.1 by assumption.

Next, note 
$$\int_{\boldsymbol{\tau}}\delta_{YXX}|((a,(b,c)),\boldsymbol{\tau})f(\boldsymbol{\tau})d(\boldsymbol{\tau})=\delta_{YXX}|(a,(b,c))$$ 
$$> \delta_{XYX}|(a,(b,c))=\int_{\boldsymbol{\tau}}\delta_{XYX}|((a,(b,c)),\boldsymbol{\tau})f(\boldsymbol{\tau})d(\boldsymbol{\tau})$$
by properties of expectations from equation \ref{expectation} and since $f(\boldsymbol{\tau})>0$ a.e. under the prior. 

Since each topology $(a,(b,c))$, $(b,(a,c))$, and $(c,(a,b))$ has a prior probability of 1/3, the maximum posterior topology will be the maximum likelihood topology. Recall from section \ref{derivation} that we can permute any site pattern probability given one topology to find a site pattern probability given another topology -- e.g.,   $\delta_{YXX}|(a,(b,c))=\delta_{XYX}|(b,(a,c))$.
So, the likelihood of trees $(a,(b,c))$ and $(b,(a,c))$ given the data are:
$$L((a,(b,c))|\boldsymbol{d}) =c_{\boldsymbol{d}} (p_{xxx})^{d_{xxx}}(p_{xxy})^{d_{xxy}}(p_{xyx})^{d_{xyx}}(p_{yxx})^{d_{yxx}}(p_{xyz})^{d_{xyz}} $$
$$L((b,(a,c))|\boldsymbol{d}) =c_{\boldsymbol{d}} (p_{xxx})^{d_{xxx}}(p_{xxy})^{d_{xxy}}(p_{xyx})^{d_{yxx}}(p_{yxx})^{d_{xyx}}(p_{xyz})^{d_{xyz}} $$
By cancelling terms in common, one can see that the likelihood ratio comes down to permuting the number of sites that follow the XYX and YXX patterns:
$$\frac{L((a,(b,c))|\boldsymbol{d})}{L((b,(a,c))|\boldsymbol{d})}= (p_{xyx})^{d_{xyx}-d_{yxx}}(p_{yxx})^{d_{yxx}-d_{xyx}}$$
Together $p_{yxx}>p_{xyx}$ and $d_{yxx}>d_{xyx}$ imply that  $\frac{L((a,(b,c))|\boldsymbol{d})}{L((b,(a,c))|\boldsymbol{d})}>1$, which implies that $\hat{S}=S$. As we have shown that $p_{yxx}>p_{xyx}$, it is sufficient to show that $d_{yxx} \rightarrow p_{yxx}$ and $d_{xyx} \rightarrow p_{xyx}$ for sufficiently large number of sites.

Let $p_{yxx,0}=\delta_{yxx}|((a,(b,c)),\boldsymbol{\tau}_0)$ and $p_{xyx,0}=\delta_{xyx}|((a,(b,c)),\boldsymbol{\tau}_0)$. From the SLLN, we have 
$$\forall \epsilon_1,\delta_1>0 \, \, \exists J_1 : \forall J>J_1 \, \, P(\frac{d_{yxx}}{J}>p_{yxx,0}-\delta_1)>1-\epsilon_1$$
$$\forall \epsilon_2,\delta_2>0 \, \, \exists J_2 : \forall J>J_2 \, \, P(\frac{d_{xyx}}{J}<p_{xyx,0}+\delta_2)>1-\epsilon_2$$
Choose $\delta_1,\delta_2$ such that $\delta_1+\delta_2<p_{yxx,0}-p_{xyx,0}$,  $\epsilon_1=\epsilon_2=\epsilon/2$, and $J_0=max\{J_1,J_2\}$ by applying the Bonferroni inequality we have that  $P(\frac{d_yxx}{J}>p_{yxx,0}-\delta_1, p_{xyx,0}+\delta_2>\frac{d_{xyx}}{J})>1-\epsilon$ and the result holds.

\end{proof}

\def\BibTeX{{\rm B\kern-.05em{\sc i\kern-.025em b}\kern-.08em
    T\kern-.1667em\lower.7ex\hbox{E}\kern-.125emX}}

\bibliography{bibfile.bib}
\bibliographystyle{plain}

\end{document}